\documentclass[11pt]{asme2ej}

\usepackage{amsmath}
\usepackage{amssymb}
\usepackage{graphicx,indentfirst,subfigure,epsfig}
\usepackage{subfigure}
\usepackage{subfigmat}
 \usepackage{varioref}
 \usepackage{wrapfig}
\usepackage{hyperref}
\usepackage{tikz}
\usepackage{cleveref}
\usepackage{booktabs}
\usepackage{etoolbox}
\usepackage{cite}
\usepackage{soul}
\usepackage{xcolor}
\usepackage[subtle]{savetrees}
\usepackage{setspace}

\usepackage{bm}

  \def\clap#1{\hbox to 0pt{\hss#1\hss}}

\providecommand{\mat}[1]{\bm{#1}}%
\renewcommand{\vec}[1]{\mathbf{#1}}

\providecommand{\mC}{\ensuremath{\mat{C}}}

\providecommand{\mI}{\ensuremath{\mat{I}}}

\providecommand{\mQ}{\ensuremath{\mat{Q}}}

\providecommand{\mS}{\ensuremath{\mat{S}}}

\providecommand{\mV}{\ensuremath{\mat{V}}}
\providecommand{\mW}{\ensuremath{\mat{W}}}
\providecommand{\mX}{\ensuremath{\mat{X}}}

\providecommand{\mLambda}{\ensuremath{\mat{\Lambda}}}

\providecommand{\va}{\ensuremath{\vec{a}}}

\providecommand{\vc}{\ensuremath{\vec{c}}}

\providecommand{\vf}{\ensuremath{\vec{f}}}

\providecommand{\vs}{\ensuremath{\vec{s}}}

\providecommand{\vu}{\ensuremath{\vec{u}}}

\providecommand{\vx}{\ensuremath{\vec{x}}}

\providecommand{\vz}{\ensuremath{\vec{z}}}

\makeatletter
\newif\if@checkentries

\makeatother

\confyear{2021}

\title{Blade Envelopes Part I: Concept and Methodology}
\author{Chun Yui Wong$^{\dagger}$\thanks{Address all correspondence to Chun Yui Wong, \texttt{cyw28@cam.ac.uk}}, Pranay Seshadri$^{\ddagger \star}$, Ashley Scillitoe$^{\star}$, 
\textbf{Andrew B. Duncan}$^{\ddagger \star}$, \textbf{Geoffrey Parks}$^{\dagger}$
    \affiliation{
    $^\dagger$Department of Engineering, University of Cambridge, U.K.\\
    $^\ddagger$Department of Mathematics, Imperial College London, U.K. \\
    $^\star$Data-Centric Engineering, The Alan Turing Institute, U.K.\\
    
    }	
}

\newcommand{\norm}[2]{\ensuremath{\left\lVert#1\right\rVert_{#2}}}
\newcommand{\mb}[1]{\mathbf{#1}}

\begin{document}

\maketitle    

\begin{abstract}
\textit{Blades manufactured through flank and point milling will likely exhibit geometric variability. Gauging the aerodynamic repercussions of such variability, prior to manufacturing a component, is challenging enough, let alone trying to predict what the amplified impact of any in-service degradation will be. While rules of thumb that govern the tolerance band can be devised based on expected boundary layer characteristics at known regions and levels of degradation, it remains a challenge to translate these insights into quantitative bounds for manufacturing. In this work, we tackle this challenge by leveraging ideas from dimension reduction to construct low-dimensional representations of aerodynamic performance metrics. These low-dimensional models can identify a subspace which contains designs that are invariant in performance---the inactive subspace. By sampling within this subspace, we design techniques for drafting manufacturing tolerances and for quantifying whether a scanned component should be used or scrapped. We introduce the blade envelope as a computational manufacturing guide for a blade that is also amenable to qualitative visualizations. In this paper, the first of two parts, we discuss its underlying concept and detail its computational methodology, assuming one is interested only in the single objective of ensuring that the loss of all manufactured blades remains constant. To demonstrate the utility of our ideas we devise a series of computational experiments with the Von Karman Institute's LS89 turbine blade.}

\end{abstract}

\section{INTRODUCTION} 
\label{sec:intro}
\begin{figure}[ht]
\centering
\includegraphics[width=0.5\linewidth]{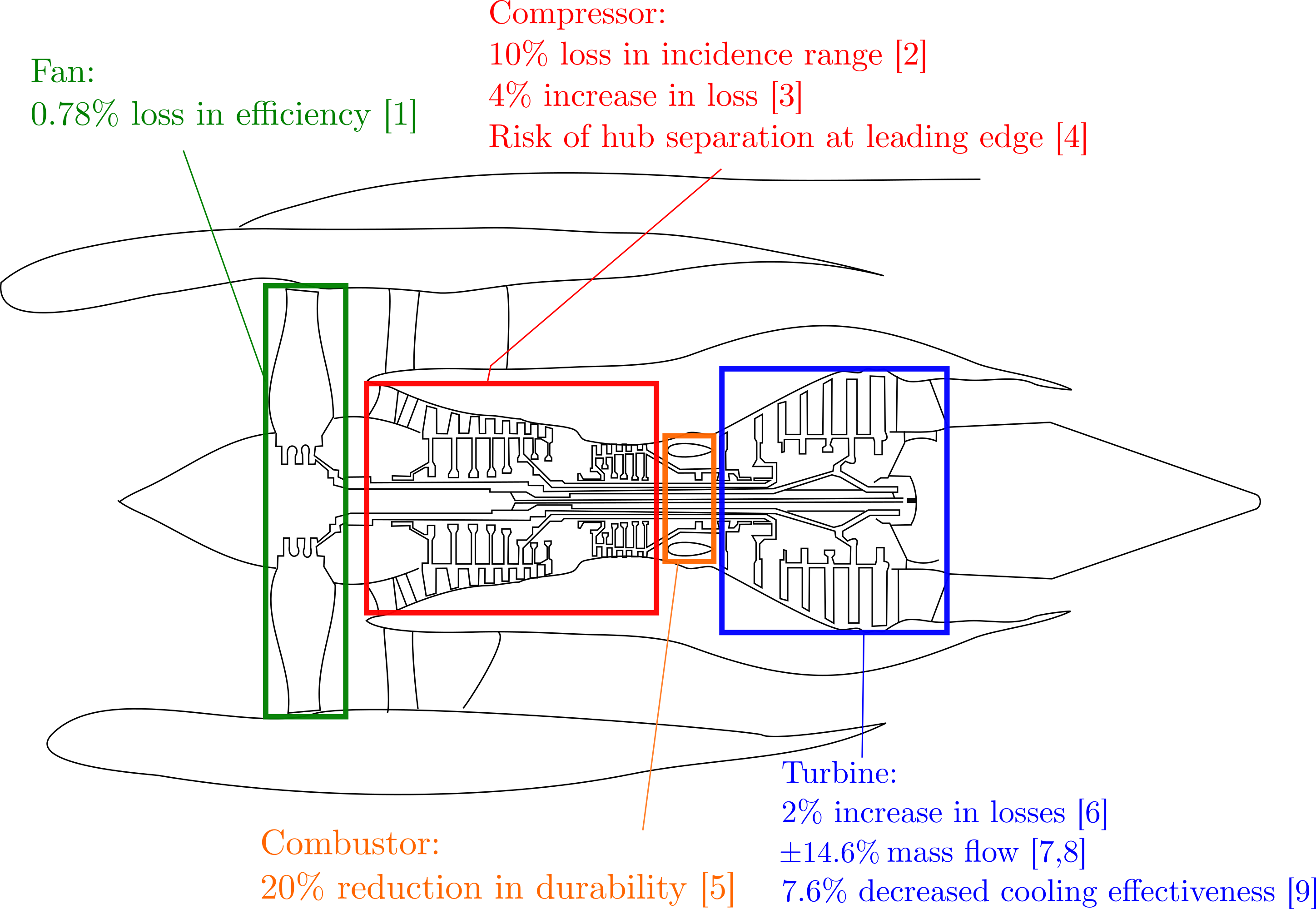}
\caption{Impacts associated with manufacturing variations in a jet engine; see  \cite{seshadri2018turbomachinery,goodhand_impact_2012,garzon_impact_2003,goodhand_impact_2014,
bradshaw_impact_2009,duffner_effects_2008,fathi_effects_2012,
wang_uncertainty_2019,shi_uncertainty_2019}.}
\label{fig:lit}
\end{figure}

Manufacturing variations and in-service degradation have a sizeable impact on the aerodynamic performance of a jet engine (see Figure~\ref{fig:lit}). 
Gauging the aerodynamic repercussions of such variability prior to manufacturing a component is challenging enough, let alone trying to predict what the amplified impact of any in-service degradation might be. In a bid to reduce losses and mitigate the risks in Figure~\ref{fig:lit}, designers today pursue a two-pronged approach. First, components are being designed to operate over a range of conditions (and uncertainties therein) via robust optimization techniques \cite{kamenik_robust_2018, seshadri_robust_2014} as well as more traditional design guides such as loss buckets---i.e., loss across a range of positive and negative incidence angles \cite{goodhand_impact_2014}. In parallel, there has been a growing research effort to assess 3D manufacturing variations and in-service degradation by optically scanning (via GOM) the manufactured blades, meshing them, and running them through a flow solver \cite{lange_principal_2012}. Both approaches, while useful in extracting aerodynamic inference, are limiting. One of the key bottlenecks is the cost of evaluating flow quantities of interest via computational fluid dynamics (CFD), as the dimensionality of the space of manufactured geometries is too large to fully explore, even with an appropriately tailored design of experiments (DoE). To reduce the dimensionality, some authors \cite{garzon_impact_2003,lange_principal_2012} use principal components analysis (PCA) to extract a few \emph{manufacturing modes}, which correspond to modes of largest manufacturing deviation observed in the scanned blades. One drawback of this approach is that the PCA model is not performance-based, i.e., the mode of greatest geometric variability need not correspond to the mode of greatest performance scatter, a point raised by Dow and Wang \cite{dow_output_2013}. Additionally, GOM scans can only be carried out on cold and manufactured components, ignoring the uncertainty in performance associated with in-service operating conditions. Finally, through neither of these paths are manufacturing engineers offered a set of pedigree rules or guides on manufacturing for an individual component, prior to actually manufacturing the component. This motivates some of the advances in this paper. We argue that challenges associated with both manufacturing variability and in-service deterioration can be adequately addressed by introducing the concept of a \emph{blade envelope}.
\begin{figure}[ht]
\centering
\includegraphics[width=0.7\linewidth]{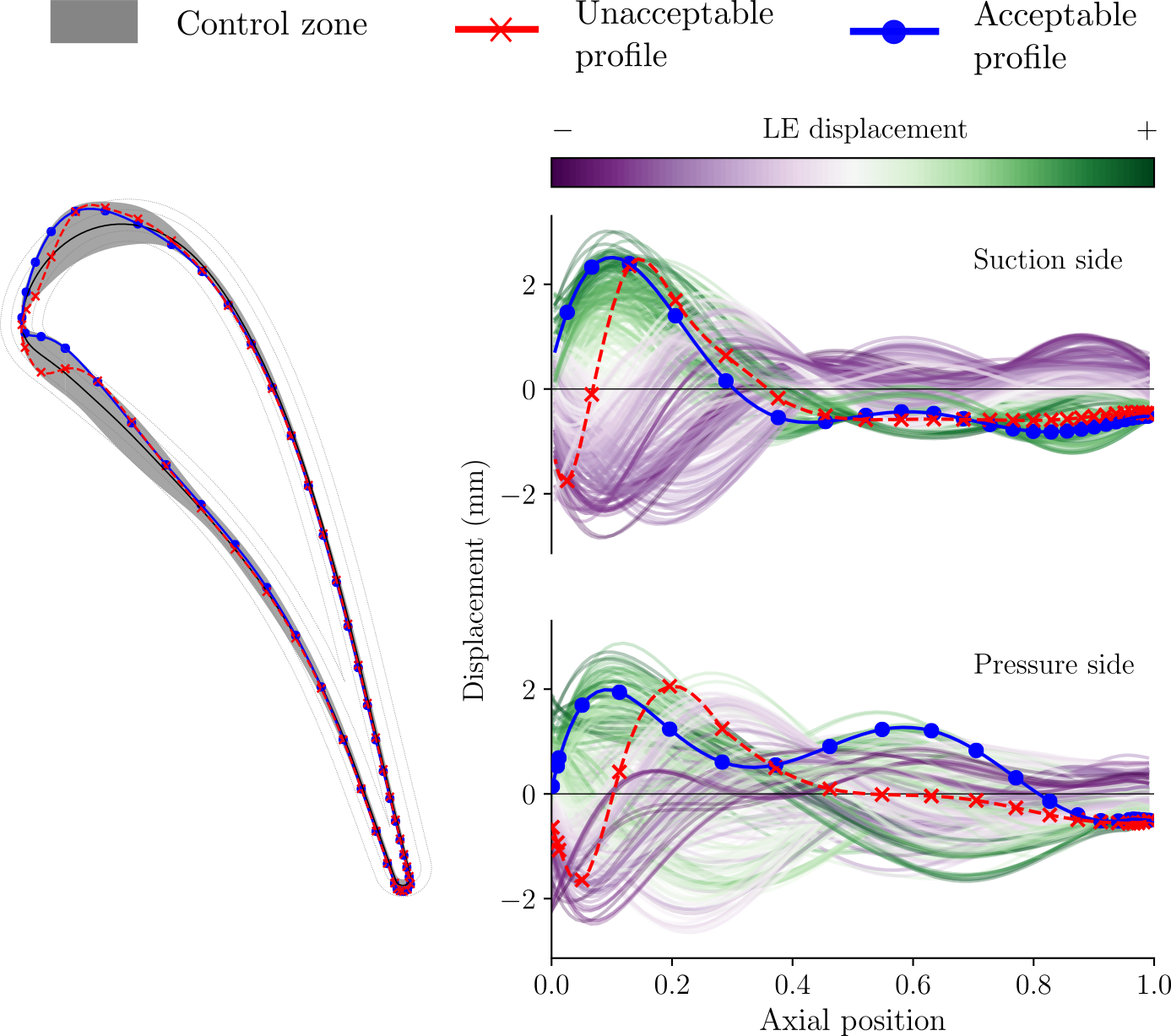}
\caption{Visual representation of a blade envelope. It comprises a control zone (grey region) and a tolerance covariance, characterized by the airfoil displacement plots color coded according to average displacement in the first 5\% of span. Dots denote CMM measurements. Note: in displacement plots, the axial distance is normalized by the axial chord. In all figures, displacements are drawn to scale.}
\label{fig:envelope}
\end{figure}

A blade envelope (see Figure~{\ref{fig:envelope}}) offers a quantitative manufacturing guide for a blade that instructs manufacturers whether a scanned component should be used or scrapped. It consists of a \emph{control zone} (shown in grey), within which a statistical distribution of perturbed geometries around a mean design $\boldsymbol{\mu}$ is defined,
\begin{equation}
BE \sim \mathrm{p}(\boldsymbol{\mu}, \boldsymbol{S}).
\end{equation}
This distribution contains geometries that result in close-to-nominal performance. Provided that a geometry lying within the control zone adheres to the \emph{tolerance covariance} $\mS$ of the distribution, it is considered to reside within the blade envelope, and is guaranteed to offer near identical performance to nominal. We offer rigorous definitions of these concepts in subsequent sections, but first offer an intuitive overview using visualizations. Key characteristics of the tolerance covariance criterion can be displayed through examination of the displacements of exemplar performance-invariant profiles within the blade envelope. Referring to Figure~{\ref{fig:envelope}}, it can be seen that leading edge (LE) displacements on both the suction and pressure sides obey a pattern---invariant profiles which are positively displaced near the LE tend also to be displaced positively in the immediate vicinity on the same side; in other words, these profiles adhere to a certain \emph{curvature constraint} dictating that the displacement cannot vary too rapidly. As an illustrative demonstration of this idea, consider two scanned profiles shown as red and blue markers in Figure~{\ref{fig:envelope}}. Each marker corresponds to airfoil profile measurements taken from a coordinate measurement machine (CMM). The measurements are interpolated using B-splines to obtain the airfoil profiles---both of which lie within the control zone. However, as can be expected, it is the profile variation within the control zone that truly dictates how adverse the change in performance will be. It is clear that the red profile, containing drastic variation in displacements over a short range, does not obey the tolerance covariance and therefore should be scrapped; indeed, this corresponds to the aerodynamic intuition that sharp geometric discontinuities are likely to cause severe performance impacts. Meanwhile, the blue geometry---having a milder curvature profile---adheres to the covariance criterion and can be used in an engine. Although we arrived at this conclusion by inspection based on rough heuristics, we stress that in practice the suitability of a profile is determined computationally using the rigorous definition of the tolerance covariance as a matrix. 

This example serves to highlight an important point---that pointwise tolerance ranges described by the control zone alone are not sufficient. The pivotal role of the tolerance covariance has profound implications on the way we think about tolerances. Consider the work of Lobato et al. \cite{lobato_united_2014} who offer a recipe for acceptance and rejection of airfoils based on the curvature variation within a chord-wise interval. Provided the curvature of the measured profile lies within a computed upper and lower tolerance band---neither of which are necessarily inferred from CFD---the profile is deemed acceptable. With blade envelopes, profile tolerance constraints on displacement and curvature are supported by computational and/or experimental aerodynamic insight. This helps avoid unnecessary scrapping due to potentially overly conservative bounds.

Beyond the binary scrap-or-use decision, blade envelopes can provide clarity on where tolerances need to be tightened and where they can be relaxed---bringing forth the possibility of not only reducing performance variability, but also reducing manufacturing scraps. This may engender transformative cost-saving manufacturing strategies for certain components, underpinned by greater predictive confidence in the tolerances. 

So, how are these tolerances obtained, and how can performance guarantees for designs manufactured within the envelope be offered? This two-part paper presents a series of computational methods to rigorously answer these questions. In this first part, we introduce blade envelopes for loss, focusing primarily on the governing methodology. In the second part \cite{wong2020bladeb}, we study how one can accommodate multiple objectives---both scalar-valued and vector-valued---when constructing such envelopes, and also elucidate their parallels to \emph{inverse design}. 
 
\section{TURBINE TEST CASE}
\label{sec:loss}
In this section, the test case used throughout this paper is introduced, along with details of its \emph{manufacturing space} and the strategy for obtaining numerical solves. However, it should be emphasized that the concept of a blade envelope and its generation is not limited to any particular profile nor restricted to any manufacturing space. 
\subsection{Geometry and design space}
We select the Von Karman Institute LS89 linear turbine cascade \cite{arts1992aero}. This transonic, highly-loaded blade serves as a rich experimental (and subsequently computational) repository with Schlieren flow visualizations, blade static pressure measurements, exit flow angle measurements, and even blade convective heat transfer values. An experimental campaign was carried  out by Arts et al. \cite{arts1992aero} at a range of different exit Mach numbers ranging from 0.70 to 1.10, exit Reynolds numbers ranging from $0.5 \times 10^5$ to $2.0 \times 10^6$ and freestream turbulence intensities of $1 \%$ to $6 \%$. The overall chord of the tested profile was approximately 67 mm with a stagger angle of $55^{\circ}$. Note that although we use a 2D axial design case for demonstration, the methodology of blade envelopes is not restricted to axial designs. Applications to centrifugal blades, and more generally to 3D turbomachinery blades will be considered in future work. However, the methodological foundation for such applications will be the same, i.e., defining the blade envelope as a statistical distribution of invariance. 

\begin{figure}
\centering
\includegraphics[width=0.25\linewidth]{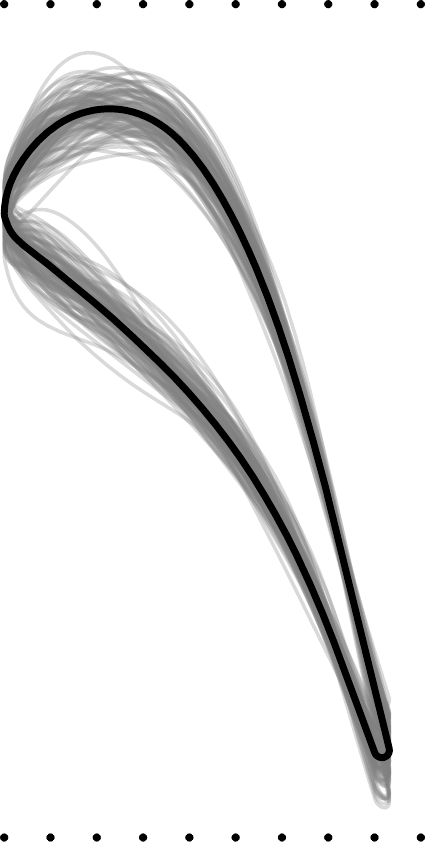}
\caption{The FFD box (black dots) with some possible deformations within the design space (grey curves) and the baseline profile (black curve).}
\label{fig:ffd}
\end{figure}

To simulate deformations arising from manufacturing variations, a large design space around the baseline profile is defined. This space, notionally denoted by $\mathcal{X}$, is parameterized by $d=20$ independent free-form deformation (FFD) design variables, \emph{scaled} to lie within the range $-1$ to $1$, such that each design vector $\vx \in [-1,1]^{d}$. Thus, one can think of $\mathcal{X}$ as being a $d$-dimensional hypercube. The FFD nodes are constrained to move only in the direction perpendicular to the inflow; some possible deformations are captured in Figure~\ref{fig:ffd}. Note that the magnitude of geometric variations imposed here grossly exaggerates the typical scales of deviations seen in real turbine blades. 

\subsection{Computational fluid dynamics}
The open-source CFD suite $SU^2$ \cite{economon_su2:_2016} is used for the mesh deformation of the various geometries as well as their subsequent numerical solves. The latter are obtained by solving the Reynolds-averaged Navier-Stokes (RANS) equations with a Spalart-Allmaras turbulence closure. No transition model is used in our work and all the boundary layers are assumed to be turbulent. A stagnation pressure driven inlet and an exit static pressure outlet were used to set the passage pressure ratio. Periodic boundary conditions were imposed along the pitch-wise direction, and the freestream turbulent viscosity ratio was set to 100. The exact pressures, temperatures and densities adopted for our studies are shown in Table~\ref{table:flow_properties_VKI}, and fall within the regime of the conditions tested during the experimental campaign. 
\begin{table}
\begin{center}
\caption{Flow properties used in this study.}
\begin{tabular}{ |l|c|c| } 
\hline
\textbf{Flow property} & \textbf{Symbol} &  \textbf{Value} \\
 \hline \hline
Inlet stagnation pressure & $p_{01}$ & $1.1 \times 10^{6}~\text{N\,m}^{-2}$ \\ 
Inlet stagnation temperature & $T_{01}$ & $592.295~\text{K}$ \\ 
Inlet density & $\rho$ & $1.2866~\text{kg\,m}^{-3}$ \\
Freestream Reynolds number & $Re$ & $6.0 \times 10^{5}$ \\
Exit static pressure & $p_{2}$ & $5.23 \times 10^{5}~\text{N\,m}^{-2}$ \\
Heat capacity ratio & $\gamma$ & 1.4 \\
 \hline
\end{tabular}
\label{table:flow_properties_VKI}
\end{center}
\end{table}
All the files and codes used to generate numerical solves can be found online at \url{https://github.com/ncywong/VKI_CFD}. We omit a detailed experimental-CFD validation here as prior comparisons of the $SU^2$ suite on this blade are available in literature \cite{vitale2015extension}.

\subsection{Key quantities of interest}
\label{sec:key_quantities}
Our interest in this paper lies in quantifying and controlling the performance variation associated with the manufacturing process. We focus on both scalar-valued (1D) quantities of interest (qoi) and vector-valued qois as functions of the profile geometry. The first performance metric we study is the stagnation pressure loss coefficient given by
\begin{equation} \label{eqn:Yp}
Y_p = \frac{p_{02} - p_{01}}{p_{02} -p_2},
\end{equation}
where $p_{02}$ is the circumferentially mass-averaged exit stagnation pressure. We also monitor the flow capacity or \emph{exit mass flow function}
\begin{equation} \label{eqn:fm}
f_m = \frac{\dot{m} \sqrt{T_{01}} }{p_{01}} \times 10^4,
\end{equation}
where $\dot{m}$ is the mass flow rate and a scale factor of $10^4$ is included for convenience. The units for $f_m$ are s$\;\cdot\;$m$\;\cdot\;$K$^{-1}$. Our final qoi is the \emph{isentropic Mach number distribution}, defined as
\begin{equation}
M(s) = \sqrt{\frac{2}{\gamma-1} \left(\left(\frac{p_{01}}{p(s)}\right)^{\frac{\gamma - 1}{\gamma}} - 1\right)},
\end{equation}
where $p(s)$ is the local static pressure at location $s$ on the surface. This profile qoi is discretized into $N=240$ discrete measurements, implying that the isentropic Mach number distribution can be defined by a \emph{vector-valued} function, 
\begin{align}
\begin{split}
\mathbf{M} &= [M(s_1),~M(s_2),~...,~M(s_N)]\\
&= [M_1,~M_2,~...,~M_N].
\end{split}
\end{align}
We create a DoE with $K$ uniformly distributed random (Monte Carlo) samples generated within our design space $\mathcal{X}$. For each design vector generated, the mesh is parametrically deformed using the FFD parameters described above and run using the $SU^2$ flow solver. The key qois are saved for each design generated, along with their corresponding design vectors.

\section{STATISTICAL METHODOLOGIES}
The input-output database from the CFD forms the bedrock of our methodology for prescribing blade envelopes. In this section, we detail some of the statistical techniques that will be applied to this database with the aim of extracting the control zone, the tolerance covariance, and for determining whether a scanned component should be used or scrapped.

\subsection{Active and inactive subspaces}
\label{sec:active}
The formulation of a blade envelope relies on the ability to identify and generate blade designs that are invariant in one or multiple objective(s). To this end, we partition the design space into two \emph{subspaces}: the \emph{active} subspace and the \emph{inactive} subspace. Here a \emph{subspace} is formed from a linear combination of all the parameters. The construction of these subspaces should satisfy the following: along the active subspace, qois vary significantly; while along the inactive subspace, qois are predominantly invariant to changes in the parameters. To clarify this breakdown, let $\mQ \in \mathbb{R}^{d \times d}$ be \emph{any} orthogonal matrix, implying that $\mQ \mQ^{T} = \mI$, where $\mI$ is the identity matrix. Now, consider the following decomposition
\begin{equation}
\begin{split}
\vx & = \mI \vx \\
& = \mQ \mQ^{T} \vx \\
& = \left[\begin{array}{cc}
\mW & \mV \end{array}\right]\left[\begin{array}{cc}
\mW & \mV \end{array}\right]^{T} \vx \\
& = \mW \mW^{T} \vx + \mV \mV^{T} \vx,
\end{split}
\label{equ:partition}
\end{equation}
where $\mW \in \mathbb{R}^{d \times r}$ and $\mV \in \mathbb{R}^{d \times (d-r)}$ are columns of $\mQ$. The matrix $\mV$ can also be obtained by determining the null space of $\mW^{T}$. The value of $r$ used in practice will be dictated by how many \emph{effective dimensions} one needs to capture the variation in the qoi(s). Ideally, rather than using \emph{any} orthogonal matrix, our goal is to identify a specific $\mQ$ that affords a partition of the design space \eqref{equ:partition} such that $r$ is as small as possible. In other words, the qoi(s) varies significantly along the directions $\mW^{T} \vx$ but are invariant to changes in $\mV^{T} \vx$. So, how do we find a $\mQ$ such that its columns admit such a dimension-reducing structure? 

Samarov \cite{samarov_exploring_1993} and Constantine \cite{constantine_active_2015} draw our attention to a covariance matrix that is the averaged outer product of the gradient
\begin{equation}
\label{eqn:grad_cov}
\mC = \int_{\mathcal{X}} \nabla_{\vx} f(\vx) \nabla_{\vx} f(\vx)^T\,\tau~d\vx,
\end{equation}
where $f$ represents the qoi, such as the loss coefficient, and $\tau = 2^{-d}$ defines a uniform distribution over the $d$-dimensional hypercube $\mathcal{X}$. The symbol $\nabla_{\vx} f \left( \vx \right) \in \mathbb{R}^{d}$ denotes the gradient of $f$ with respect to its shape parameters $\vx$. As $\mC$ is a symmetric matrix, one can write its eigendecomposition as
\begin{equation}
\mC = \mQ \mLambda \mQ^T
\end{equation}
where $\mQ$ is an orthogonal matrix of $\mC$'s eigenvectors and $\mLambda$ a diagonal matrix of its eigenvalues. Indeed, \emph{this is} the $\mQ$ we shall use to partition our design space as per \eqref{equ:partition}. The value of $r$ is set by the spectrum of eigenvalues in $\mLambda$. We consider the separation
\begin{equation}\label{eqn:eigenval}
\mC = [\mW \enskip \mV] \begin{bmatrix} 
\mLambda_1 & \mathbf{0}\\
\mathbf{0} &\mLambda_2
\end{bmatrix}
\begin{bmatrix}
\mW^T\\
\mV^T
\end{bmatrix},
\end{equation} 
where $\mLambda_1$ is a diagonal matrix containing the largest eigenvalues, and $\mLambda_2$ the smallest eigenvalues, both sorted in descending order. This partition should be chosen such that there is a large gap between the last eigenvalue of $\mLambda_1$ and the first eigenvalue of $\mLambda_2$ \cite{constantine_active_2015}. Thus, this partitioning of $\mQ$ yields the active subspace matrix $\mW$ and the inactive subspace matrix $\mV$.

\subsection{Sparse polynomials for gradient estimation}
\label{sec:sparse}
\begin{figure}[h]
\begin{center}
\begin{subfigmatrix}{2}
\subfigure[]{\includegraphics[width=0.3\textwidth]{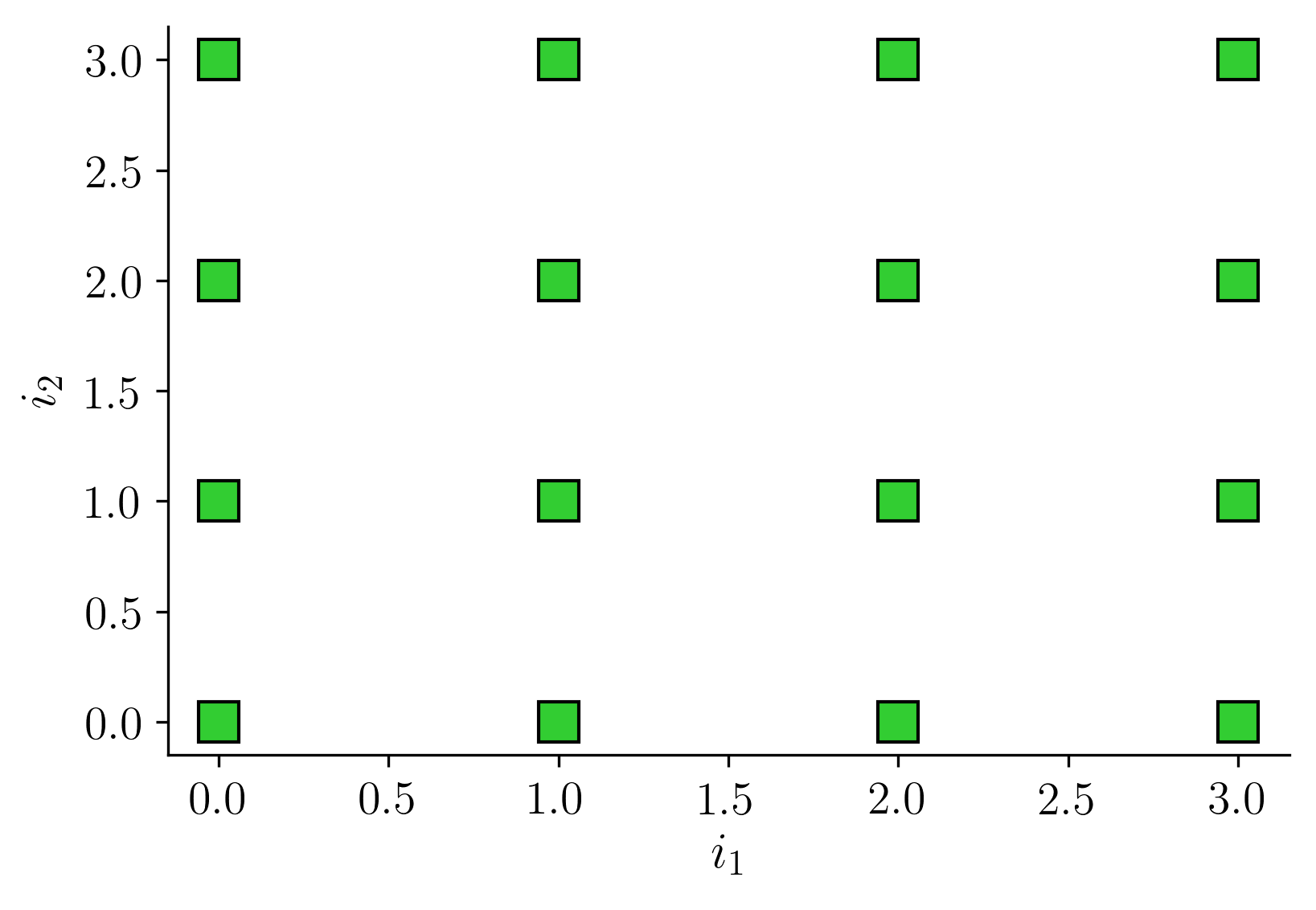}}
\subfigure[]{\includegraphics[width=0.3\textwidth]{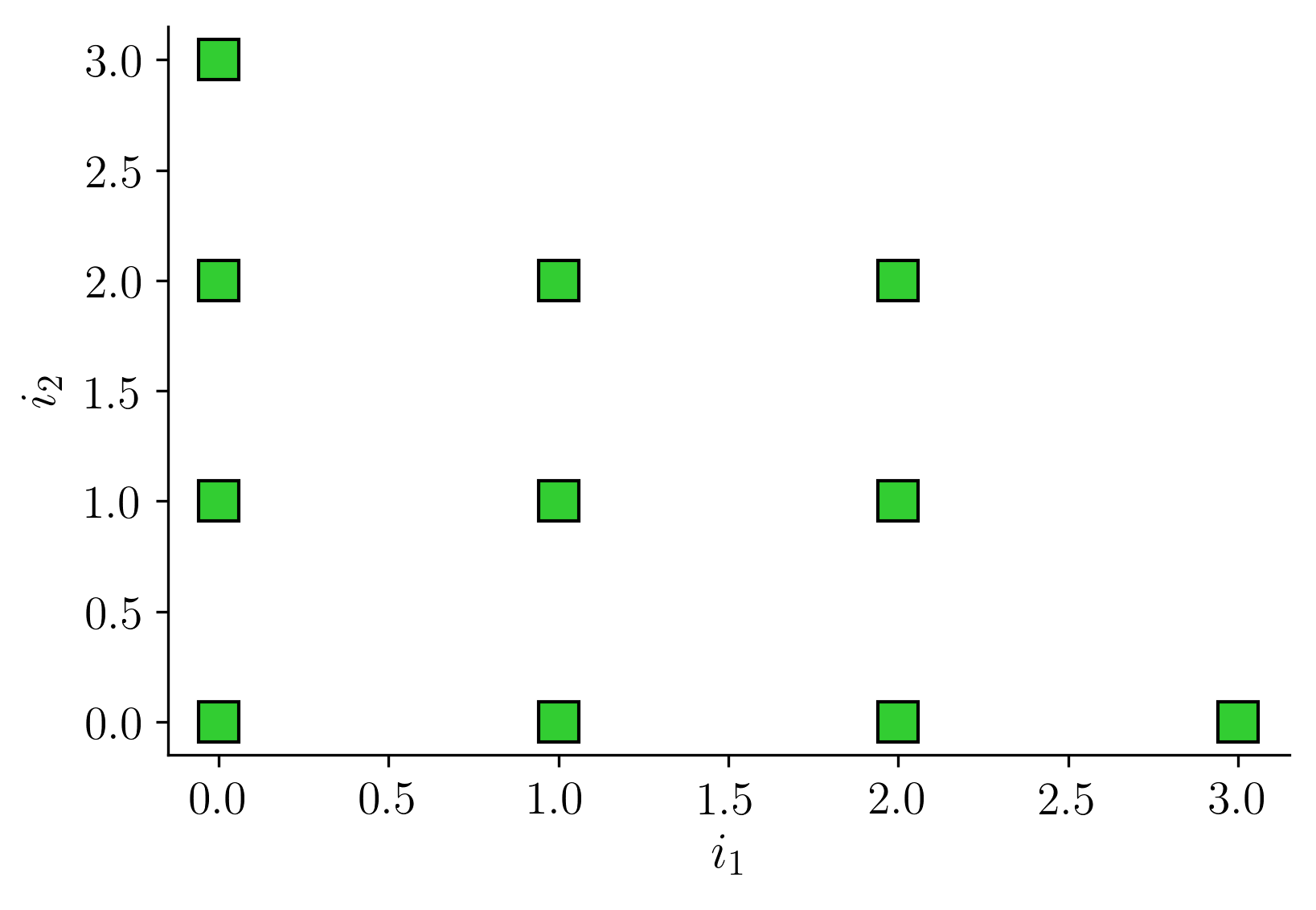}}
\subfigure[]{\includegraphics[width=0.3\textwidth]{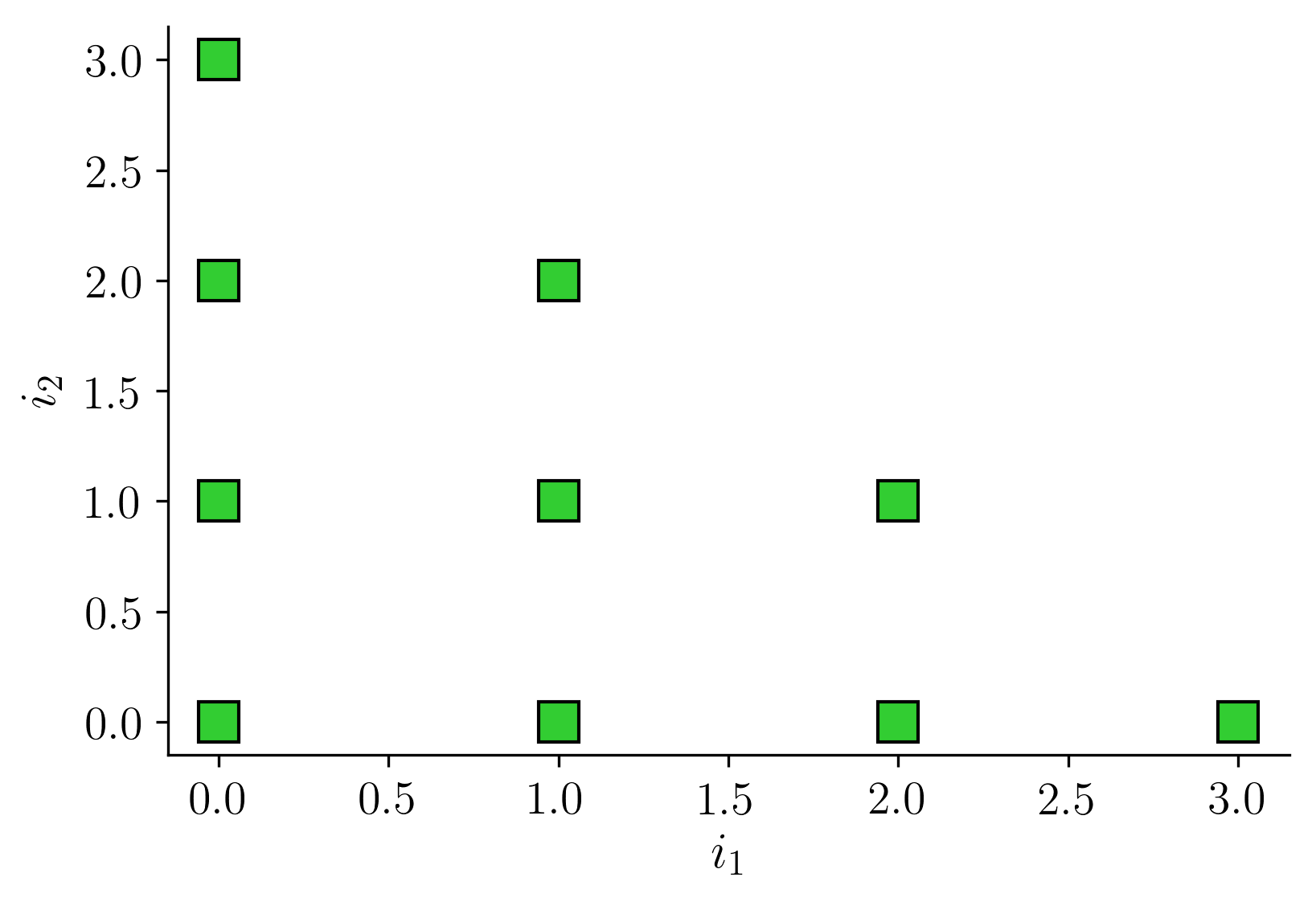}}
\subfigure[]{\includegraphics[width=0.3\textwidth]{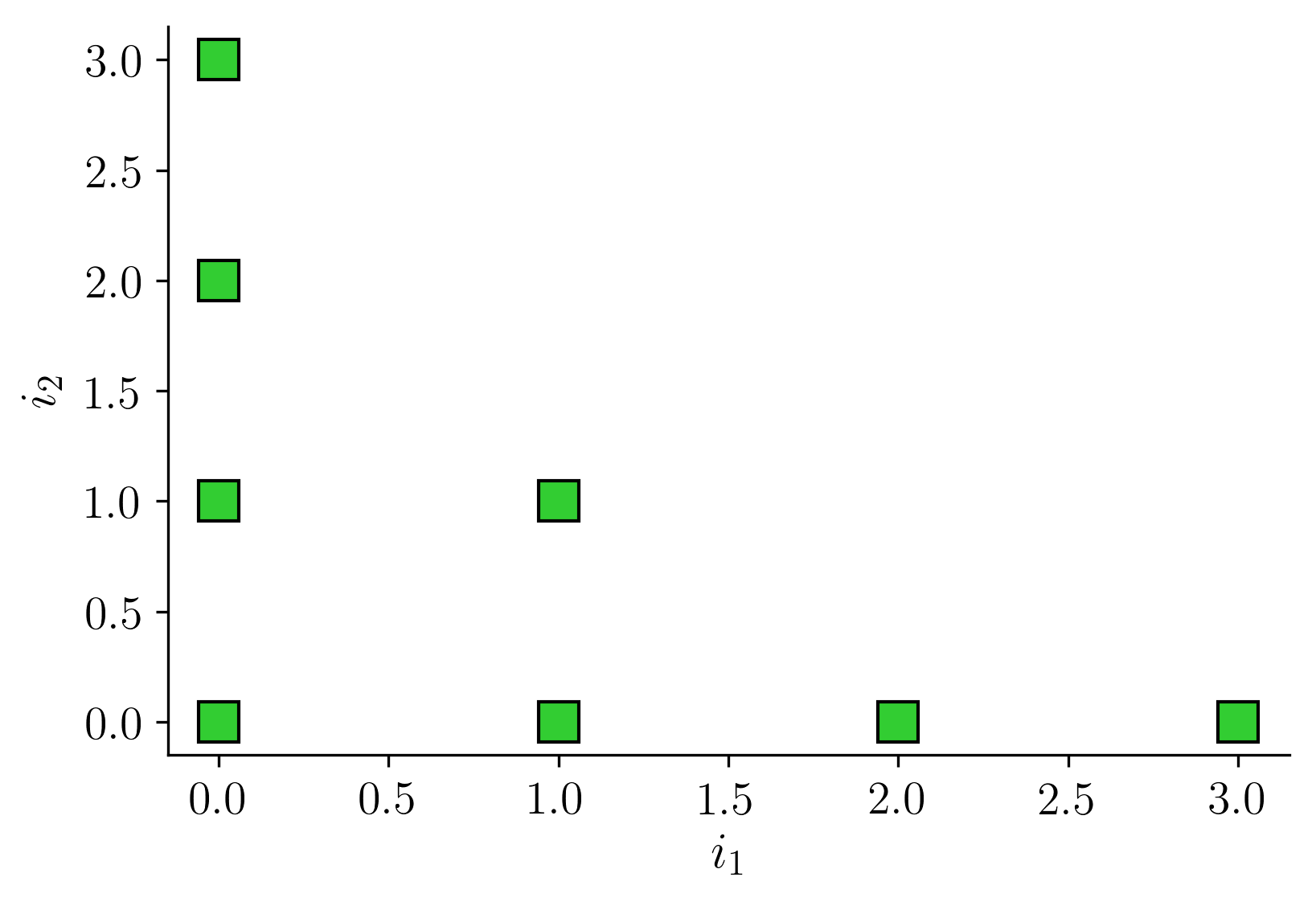}}
\end{subfigmatrix}
\caption{Standard multi-index set $\mathcal{I}$ options: (a) tensorial; (b) Euclidean degree; (c) total order; and (d) hyperbolic basis for $d=2$ and $p=3$.}
\label{fig:mulit_index_set}
\end{center}
\end{figure}

One challenge with the subspace idea detailed above is the necessity for gradient evaluations of the qoi (see \eqref{eqn:grad_cov}). Gradient evaluations can be efficiently computed using the adjoint method. However, automatic differentiation tools may not always be available, particularly in older industrial codes. In this paper, we estimate the gradients through a global polynomial approximation over the design space
\begin{equation} \label{eqn:polymodel}
f\left( \vx \right)   \approx  \sum_{\boldsymbol{i} \in \mathcal{I}}^{O} a_{\boldsymbol{i}}  \boldsymbol{\psi}_{\boldsymbol{i}}(\vx)
\end{equation}
where the polynomial coefficients are denoted by $a_{\boldsymbol{i}}$ and the basis terms are $\boldsymbol{\psi}_{\boldsymbol{i}} \left( \vx \right)$. These basis terms are formed from the product of univariate orthogonal polynomials and the \emph{multi-index set} $\mathcal{I}$ in \eqref{eqn:polymodel} dictates which $O$ permutations of univariate polynomials of varying degrees are present. More specifically, one can think of each multivariate polynomial as 
\begin{equation}
\boldsymbol{\psi}_{\boldsymbol{i}} (\vx)= \prod_{j=1}^{d} \psi_{i_{j}}^{\left(j\right) } \left(x^{\left(j\right)} \right), \; \; \; \; \; \text{where} \; \; \; \; \; \boldsymbol{i} = \left(i_1, \ldots, i_d \right).
\end{equation}
The superscript $\left(j\right)$ denotes the univariate polynomial along dimension $j$, which is only a function of the univariate parameter $x^{\left( j \right)}$. The degree for each univariate polynomial $\psi_{i_{j}}^{\left(j\right) }$ is specified in the \emph{multi-index} $\boldsymbol{i}$. All the different multi-indices lie in the finite multi-index set $\mathcal{I}$, mentioned above. To qualitatively draw parallels between different multi-index sets, Figure~\ref{fig:mulit_index_set} shows typical isotropic multi-index permutations for $d=2$ where the highest degree is restricted to $p=3$. The number of squares in each of these subfigures corresponds to $O$, i.e., the number of coefficients that need to be computed.
\patchcmd{\subfigmatrix}{\hfill}{\hspace{0.8cm}}{}{}
Selecting the appropriate multi-index set is an important aspect of approximating multivariate functions via polynomials (see \cite{seshadri2017effectively}). Decisions should be made depending on whether one would expect higher order interactions among the mixed terms or whether most of the large coefficients are associated with lower order principal terms. The total order basis, shown in Figure~\ref{fig:mulit_index_set}(c), is often a good compromise and ameliorates problems related to the exponential scaling in the number of associated coefficients with increasing $d$---an artifact of the tensorial basis in (a). The number of coefficients for a total order basis is given by
\begin{equation}
O =  \left(\begin{array}{c}
p+d\\
p
\end{array}\right) = \frac{ \left(p + d\right)! }{  p! \; d! }.
\label{equ:total_order}
\end{equation}
Now, to understand the relationship between the number of coefficients $O$ and the number of CFD evaluations $K$ required to obtain an approximation, let the input-output qoi database be given by
\begin{equation}
\mX=\left[\begin{array}{c}
\vx_{1}^{T}\\
\vdots\\
\vx_{K}^{T}
\end{array}\right], \; \; \; \; \; \vf=\left[\begin{array}{c}
f_{1}\\
\vdots\\
f_{K}
\end{array}\right],
\end{equation}
where $\mX \in \mathbb{R}^{K \times d}$ is the matrix of inputs and $\vf$ is a vector of the qois across the $K$ CFD evaluations. We can write the coefficients and discretized basis terms in \eqref{eqn:polymodel} as
\begin{equation}
\va=\left[\begin{array}{c}
a_{1}\\
\vdots\\
a_{o}
\end{array}\right],  \; \; \; \; \; \; \boldsymbol{\Psi}=\left[\begin{array}{ccc}
\boldsymbol{\psi}_{1} \left( \vx_{1}\right) & \ldots & \boldsymbol{\psi}_{O}\left(\vx_{1}\right)\\
\vdots & \ddots & \vdots\\
\boldsymbol{\psi}_{1}\left(\vx_{K}\right) & \cdots & \boldsymbol{\psi}_{O}\left(\vx_{K}\right)
\end{array}\right].
\end{equation}
Once the coefficients $\va$ are known, gradient approximations can be readily computed via
\begin{equation}
\nabla_{\vx} f \left( \vx \right) \approx \sum_{\boldsymbol{i} \in \mathcal{I}}^{O} a_i \nabla_{\vx} \boldsymbol{\psi}_{\boldsymbol{i}}(\vx).
\end{equation}
This idea of extracting gradients for estimating $\mC$ using a polynomial approximation was previously considered in \cite{seshadri2018turbomachinery} where the authors constructed a global least squares quadratic fit. The trouble with a least squares approach is that it requires $K \geq O$, i.e., we need as many  CFD evaluations as basis terms (and ideally more), which can be prohibitive in high-dimensional problems. 

To abate this computational cost, we propose the use of compressed sensing for estimating the coefficients; for an introductory background see \cite{bryan2013making}. This involves the solution to the optimization problem
\begin{equation}
\begin{split}
\underset{\va}{ \text{minimize} } \; & \;  \left\Vert \va \right\Vert _{1} \\
\text{subject to} \; & \; \norm{\boldsymbol{\Psi} \va - \vf}{2} \leq \epsilon,
\end{split}
\label{equ:cs}
\end{equation}
where the notation $\left\Vert \cdot \right\Vert _{1}$ denotes the $L_1$ norm, i.e., the sum of the absolute values of the arguments; $\left\Vert \cdot \right\Vert _{2}$ denotes the familiar $L_2$ or Euclidean norm, i.e., the square root of the sum of the squares of the argument. Here the small positive constant $\epsilon$ is used to account for the mis-match between the polynomial approximation and the true function evaluations $\vf$. 

Central to the compressed sensing paradigm is the notion of \emph{sparsity} in the coefficients. In other words, to solve \eqref{equ:cs} we assume that a fraction of the coefficients are close to zero, although we do not need to know precisely which ones are. This enables us to solve for $\va$ even when $\boldsymbol{\Psi}$ has more columns than rows, i.e., we have more unknown coefficients than CFD evaluations. More broadly, the use of compressed sensing for polynomial approximations has received considerable attention in literature \cite{peng2014weighted}, particularly for applications focused on uncertainty quantification. 

To summarize, the salient advantage of solving \eqref{equ:cs} is the reduction in the number of CFD evaluations required. We can extract solutions for $\va$ even when $K \ll O$, provided that $\va$ is sparse. This is very useful as we would like to reduce the number of function evaluations required for estimating the gradients and subsequently the active and inactive subspaces.

\subsection{Sampling from the inactive subspace}
\label{sec:sample}
\begin{figure}[ht]
\centering
\includegraphics[width=0.7\linewidth]{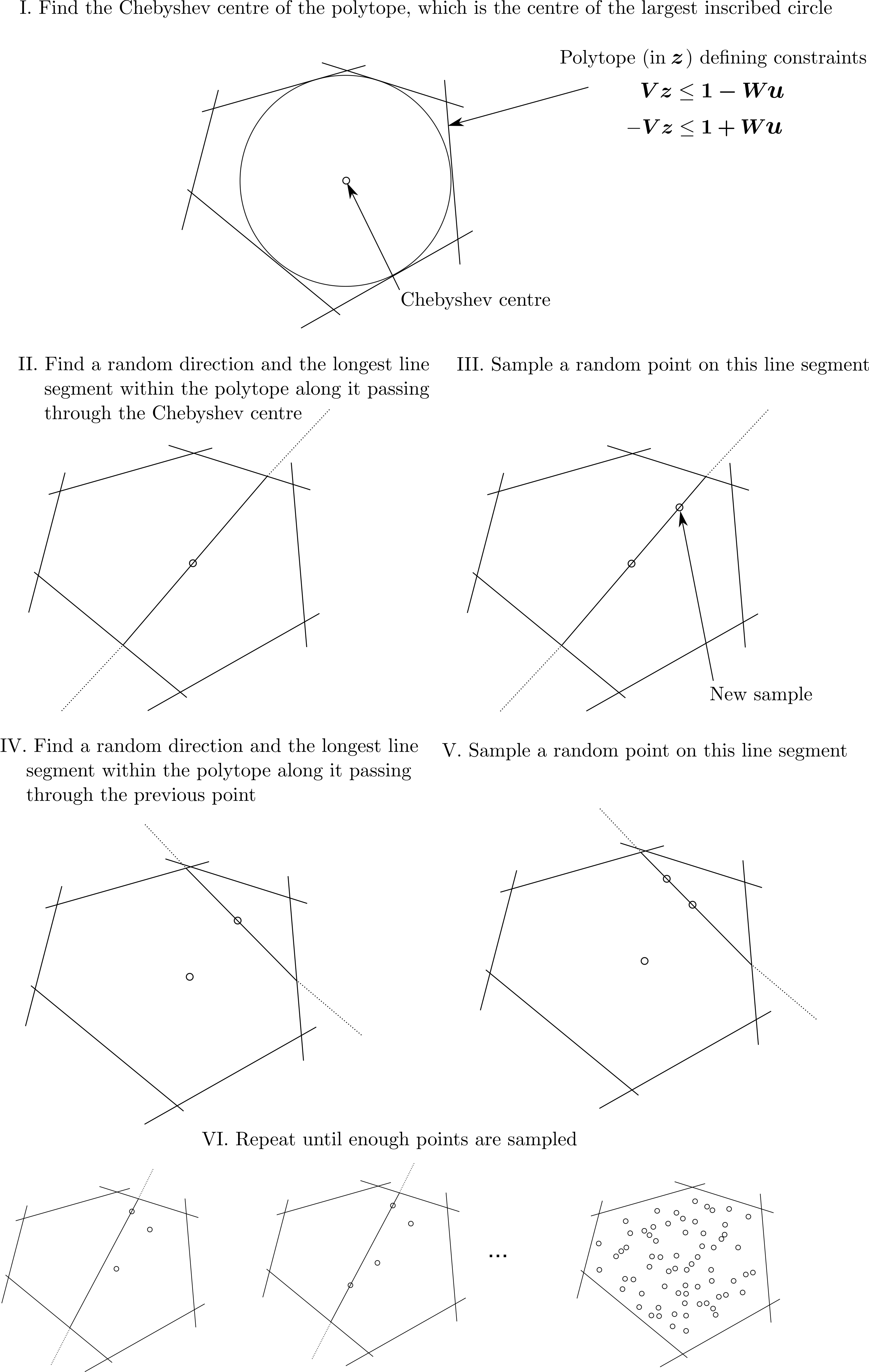}
\caption{Schematic showing the hit-and-run sampling method for generating samples in the inactive subspace.}
\label{fig:inactive}
\end{figure}

Equipped with a covariance matrix $\mC$ that facilitates dimension reduction, our goal is to generate blade profiles that will be provably invariant. One way to do this is to generate samples from the inactive subspace $\mV$ for a constrained value along the active subspace $\mW$. Building upon \eqref{equ:partition}, we write
\begin{equation}
\begin{split}
\vx & =  \mW \mW^{T} \vx + \mV \mV^{T} \vx \\
&  = \mW \vu + \mV \vz,
\end{split}
\label{equ:partition2}
\end{equation}
where $\vu := \mW^T \vx$ is the active coordinate and $\vz := \mV^T \vx$ the inactive coordinate. Our task is to find points $\vx$ with a fixed $\vu$ but different $\vz$, while obeying the constraint $-\mb{1} \leq \vx \leq \mb{1}$. The bounds imply that
\begin{align}
\begin{split}
\mV \vz \leq \mb{1} - \mW \vu,\\
-\mV \vz \leq \mb{1}+\mW\vu.
\label{equ:hitandrunbounds}
\end{split}
\end{align}
These linear constraints define a \emph{polytope} in the inactive subspace (see Figure~\ref{fig:inactive} I). 

To generate samples from this polytope, we adopt the hit-and-run method \cite{smith_hit-and-run_1996} based on the implementation by Constantine et al. \cite{constantine_python_2016}. In this method, one first identifies a feasible point by locating the Chebyshev centre of the polytope\footnote{The Chebyshev centre is the centre of the largest hypersphere that can fit within the polytope (see page 418 in \cite{boyd2004convex}).} at the specified active coordinate $\vu$, by solving a linear program. Then, starting from the Chebyshev centre, a random direction is selected and, in this direction, the longest line segment that lies within the polytope (and goes through the centre) is identified. Following this, a point on this line segment is selected at random yielding the first sample. This procedure is repeated by starting with a new random direction and identifying the longest line segment along that line. These steps are captured in Figure~\ref{fig:inactive} II to VI.

\subsection{Machine learning blade envelopes}
\label{sec:distance}
The statistical methodologies presented thus far set the stage for generating both the \emph{control zone} and the \emph{tolerance covariance}---the two key elements of the blade envelope. To formalize these two, let $H$ be the number of inactive samples generated via the hit-and-run method for the nominal value of $\vu$ at the origin---corresponding to the undeformed geometry. Furthermore, let $\vs \in \mathbb{R}^{N}$ denote the vertical coordinates of each airfoil corresponding to these inactive samples, where we assume that the horizontal coordinates are the same over all the samples. In other words, the airfoil coordinates $\left\{ \vs_{1}, \vs_{2}, \ldots, \vs_{H}\right\}$ represent designs that are invariant in the selected qoi(s). The bounds of the control zone are then specified by the upper and lower limits $\vc_u, \vc_l \in \mathbb{R}^{N}$ where
\begin{align}
\begin{split}
\vc_u = \; \; \text{maximum} \; \; \left\{ \vs_{1}, \vs_{2},\ldots, \vs_{H}\right\},\\
\vc_l = \; \; \text{minimum} \; \; \left\{ \vs_{1}, \vs_{2},\ldots, \vs_{H}\right\},
\end{split}
\end{align}
where the maximum and minimum are taken coordinate-wise. The control zone by itself is very useful to offer some insight into where tolerances need to be tightened and where they may potentially be relaxed. 

The \emph{tolerance covariance} matrix quantitatively captures the geometric characteristics of invariant profiles in the control zone. As demonstrated in Figure~{\ref{fig:envelope}}, it can be notionally interpreted as a constraint on the curvature variation on both the suction and pressure sides, though in practice one should adopt a computational characterization in lieu of deriving qualitative heuristics based on inspection. Formally, we define the tolerance covariance as the matrix $\mS \in \mathbb{R}^{N \times N}$ where each element is given by
\begin{equation} \label{eqn:tol_cov_matrix}
\mS_{ij} = \text{cov} \left( s^{(i)},~ s^{(j)} \right),
\end{equation}
where the superscript $(i)$ denotes the $i$-th discrete measurement coordinate. The diagonal entries of this matrix refer to pointwise variance, related to the  average variation of profiles at each measurement location, while the off-diagonal entries encode the pairwise correlation between the deviations at different locations. This has a close relation with the typical curvature profile of invariant geometries.

As stated earlier, when a manufactured component is scanned, we wish to determine whether it complies with a computed blade envelope for the design in question. This necessitates a machine learning paradigm where we can map the CMM readings of a blade onto its blade envelope and distill a binary scrap-or-use output. We will assume that, although the number of CMM points may be less than $N$, one can linearly interpolate them to estimate the vertical displacements (or surface normal displacements) at the same horizontal coordinates (or chord-wise locations) as the blade envelope. Alternatively, one can do the opposite and downsample the coordinates associated with the blade profile, which are obtained solely from CFD. The resolution of the CMM measurement and fidelity of the CFD simulation limits the scale of the smallest features that can be captured by the blade envelope.

\begin{figure}[ht]
\centering
\includegraphics[width=0.5\linewidth]{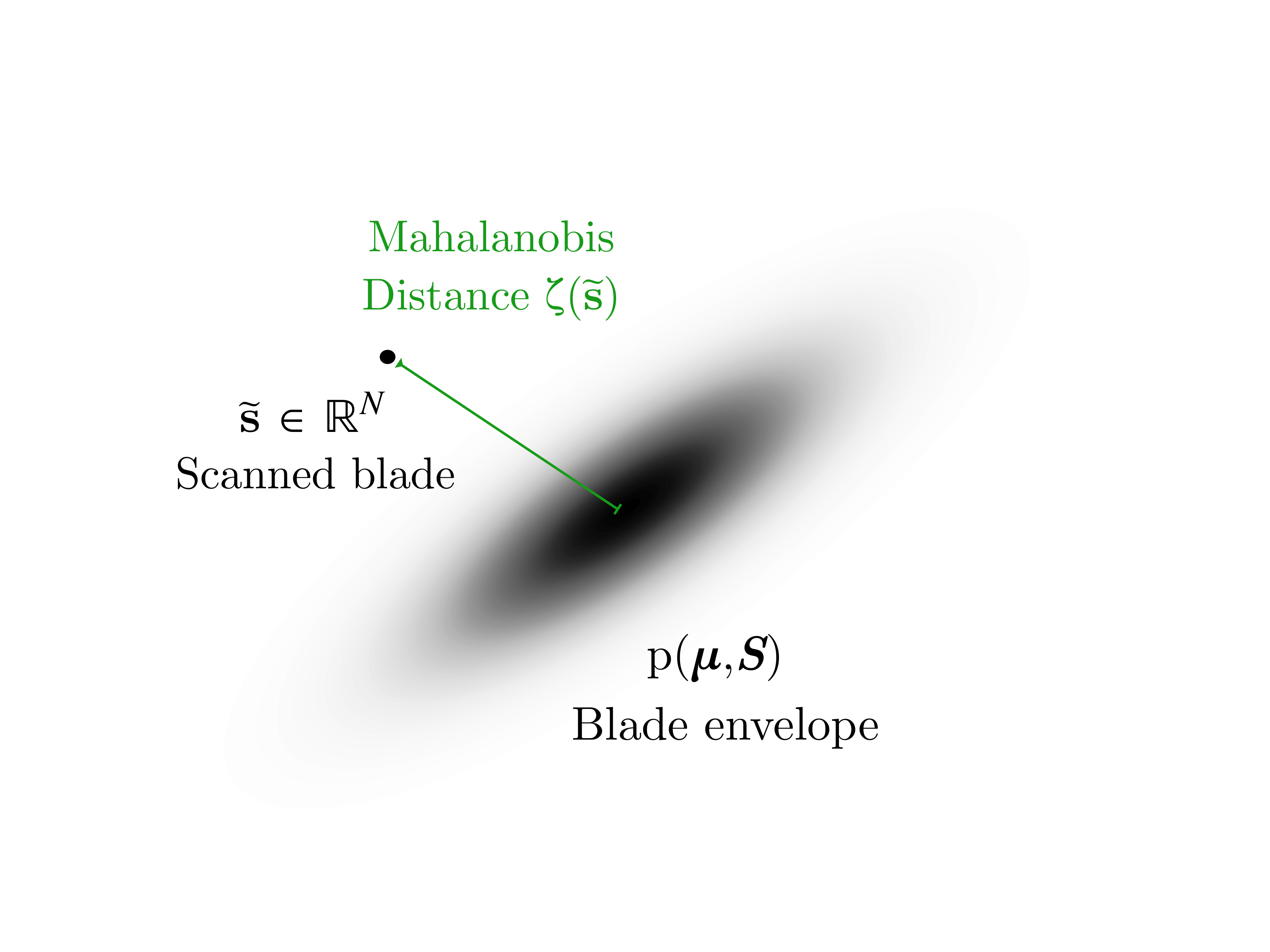}
\caption{Illustrating the Mahalanobis distance metric.}
\label{fig:maha_dist_fig}
\end{figure}

We compute the Mahalanobis distance between the scanned blade and the distribution associated with the blade envelope. The Mahalanobis distance is a statistical measure of the distance between a point $\tilde{\vs}$ and a distribution (see Figure {\ref{fig:maha_dist_fig})}, 
\begin{equation}
\zeta \left( \tilde{\vs} \right) = \sqrt{ \left( \tilde{\vs} - \boldsymbol{\mu} \right)^T \mS^{-1}  \left( \tilde{\vs} - \boldsymbol{\mu} \right)  },
\label{equ:distance}
\end{equation}
which is a Euclidean distance weighted by the inverse of the covariance matrix $\mS$. In our case, the blade envelope is a distribution defined over the space of geometries parameterized by $N$ coordinates and characterized by the mean profile $\boldsymbol{\mu}$ and the tolerance covariance $\mS$; a sample profile is a point in this space, and we wish to calculate its distance away from the blade envelope. Although we can generate infinitely many airfoil profiles (see Figure~\ref{fig:envelope}), we do not know the true distribution from which these samples are obtained. However, we can still estimate the \emph{ensemble mean}
\begin{equation}
\boldsymbol{\mu} = \frac{1}{H} \sum_{i=1}^{H} \vs_i.
\end{equation}
From this, the Mahalanobis distance is calculated according to \eqref{equ:distance} by substituting the tolerance covariance. If the distance is very small, then the new profile $\tilde{\vs}$ is said to lie within the blade envelope and is thus deemed acceptable. On the other hand, if the distance is large, the new profile clearly lies far away from the distribution associated with the blade envelope and therefore should not be used. 

We can convert this distance---a relative measure of acceptance and rejection---to an absolute one via a logistic function
\begin{equation}
g \left( \tilde{\vs} \right) = \frac{\beta_1}{1 + \text{exp}  \left\{  -\beta_2 \left( \zeta \left( \tilde{\vs} \right) - \beta_3 \right)  \right\}   }
\label{equ:logistic}
\end{equation}
where model parameters $\beta_1, \beta_2$ and $\beta_3$ have to be determined by a training-testing criterion. Generating training and testing samples requires no additional CFD evaluations, and the parameters themselves can be estimated via standard gradient-based optimization. Once determined for a specific blade, expression \eqref{equ:logistic} can be used on the factory floor to render a scrap-or-use judgment for a manufactured blade. 

We remark that here that one alternative to training the logistic function in \eqref{equ:logistic} is to estimate the critical values of the Mahalanobis distance, given many scanned blades. As these distance metrics yield a chi-squared distribution, one can assign a \emph{significance level} to the distribution to deliver a scrap-or-use verdict on all the blades scanned.

\subsection{Summarizing the key steps}
\begin{figure}[ht]
\centering
\includegraphics[width=0.5\linewidth]{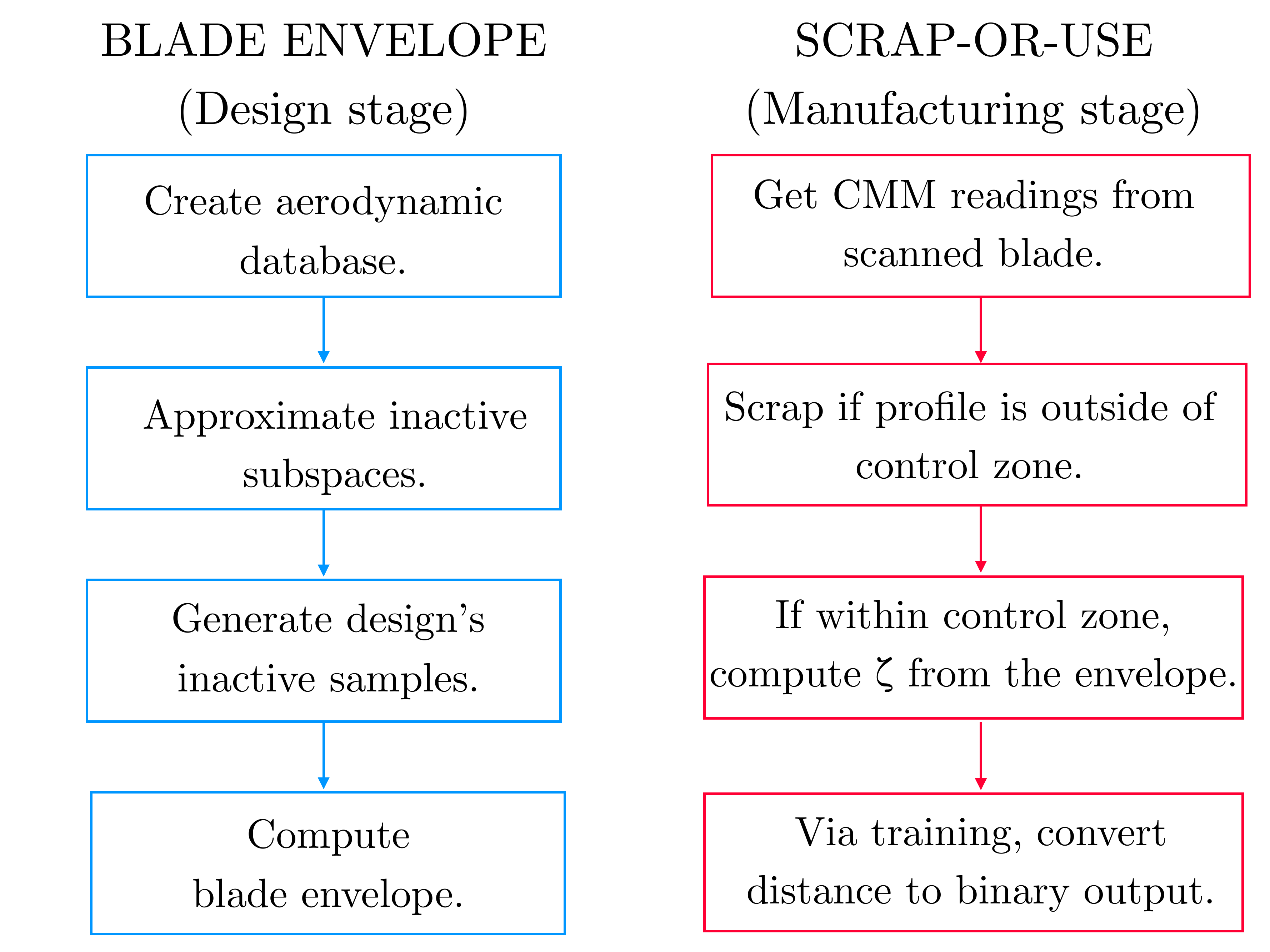}
\caption{Flowchart summarizing the key steps involved in generating a blade envelope and querying it to ascertain whether a scanned component should be used.}
\label{fig:schematic}
\end{figure}

We conclude this section with a summary of the main techniques involved in generating a blade envelope and exploiting it to query whether a scanned blade should be used; these steps are captured in Figure~\ref{fig:schematic}. 

For generating a blade envelope, the first requirement is a well-sampled aerodynamic database that comprises a collection of geometries around the baseline profile and the output performance of each geometry (\ref{sec:key_quantities}). Using this input-output database, we can approximate the inactive subspaces (\ref{sec:active} and \ref{sec:sparse}) and generate sample designs that will admit the same performance, but have very different geometries (\ref{sec:sample}). By following this workflow, we can generate the blade envelope, shown in Figure~\ref{fig:envelope}. 

The mean and covariance of the \emph{inactive samples} associated with this blade envelope can then be used to ascertain whether a scanned blade's CMM coordinates are likely to lie within this envelope (\ref{sec:distance}). A logistic function can be trained to yield a binary result that can be deployed in various blade manufacturing sites.

\section{BLADE ENVELOPES FOR 1D QUANTITIES}
In this section, we use the techniques discussed above to identify and sample from the inactive subspace associated with loss. Our focus is to deliver a scrap-or-use decision for blade profiles that could have been manufactured for the LS89 profile, if near-design loss was the only performance consideration.

\subsection{Inactive subspace for loss}
\begin{figure}[h]
\begin{center}
\begin{subfigmatrix}{1}
\subfigure[]{\includegraphics[width=0.4\linewidth]{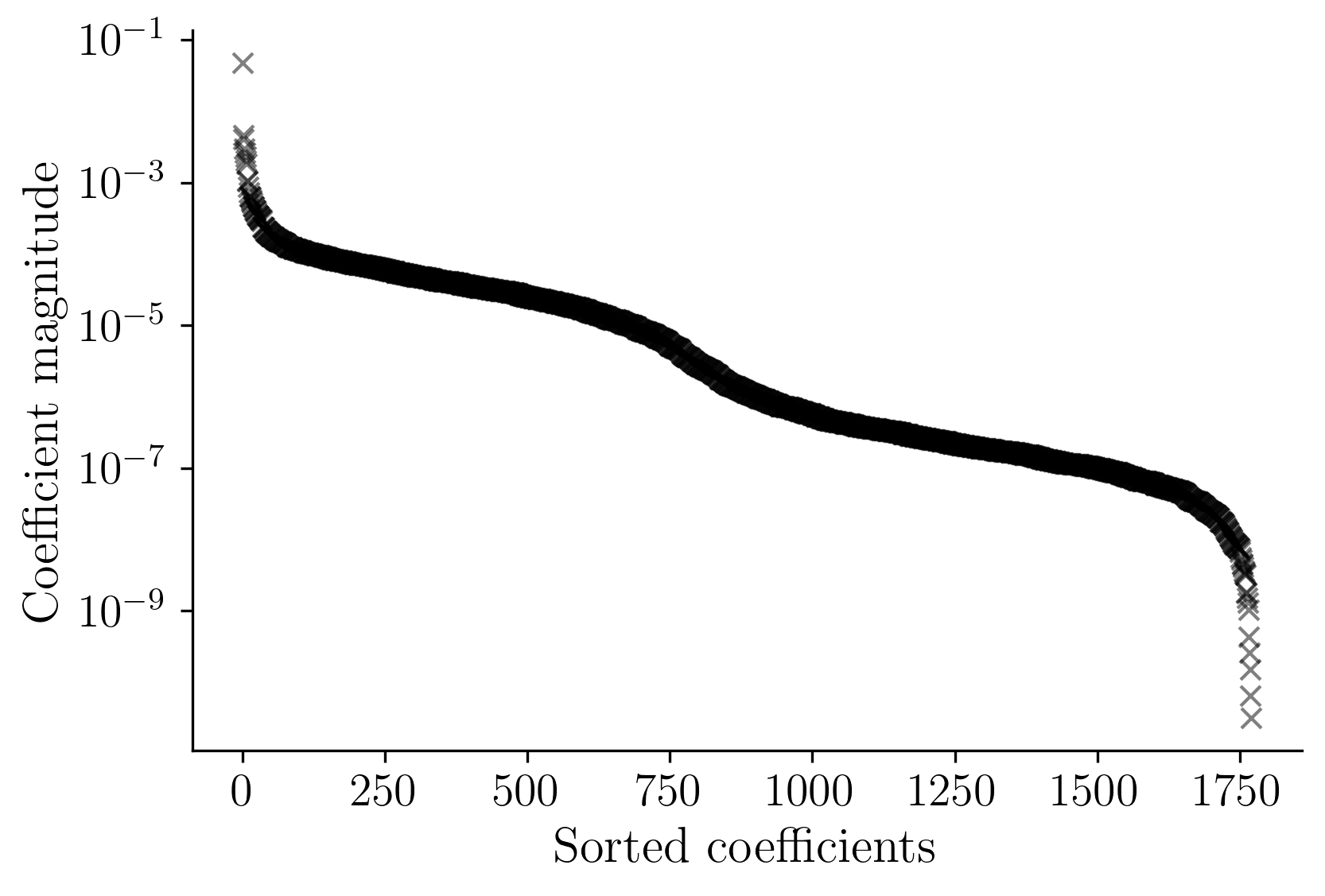}}
\subfigure[]{\includegraphics[width=0.4\linewidth]{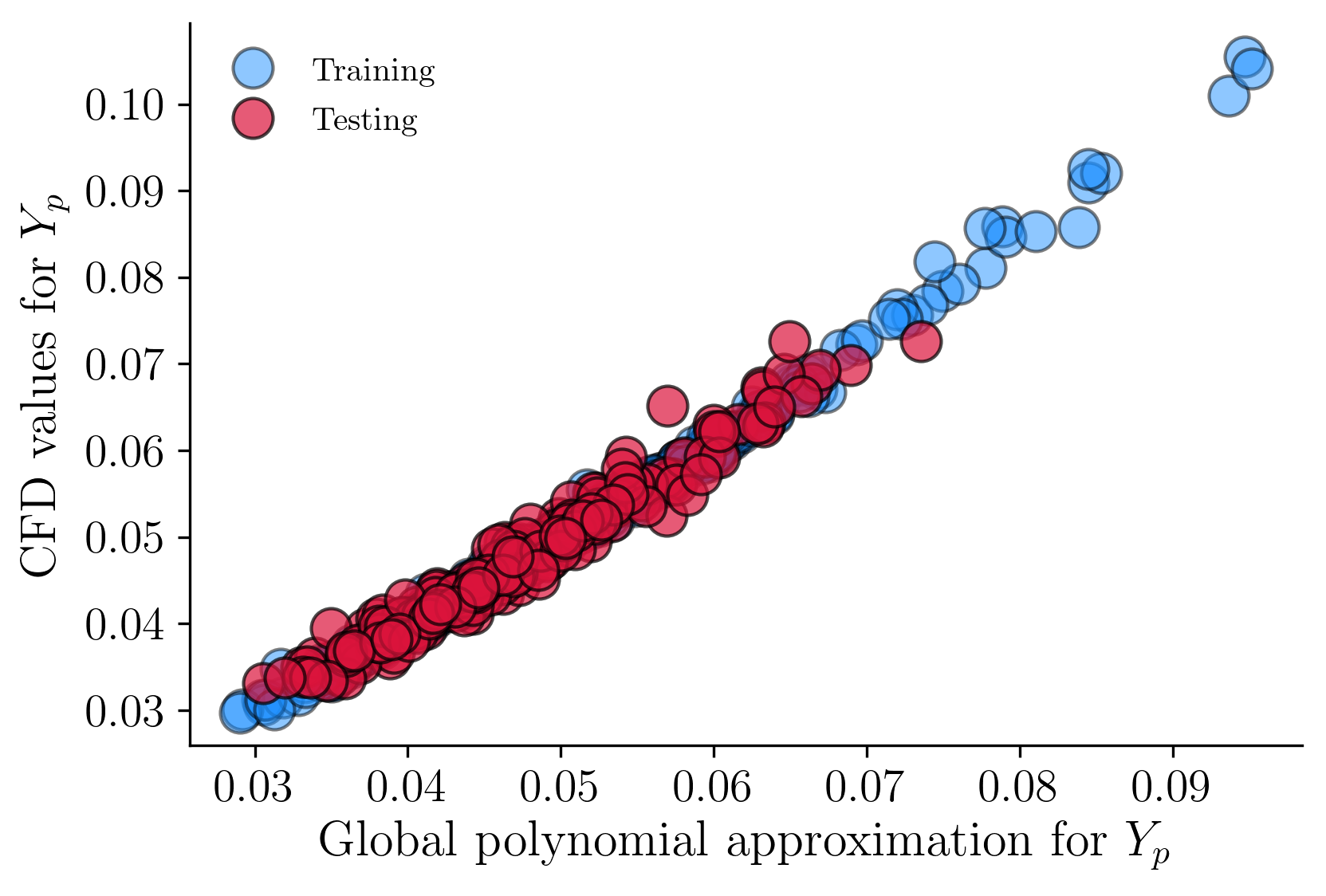}}
\end{subfigmatrix}
\caption{Global sparse polynomial approximation: (a) sorted coefficients; (b) validation of the model using the testing data.}
\label{fig:global_poly}
\end{center}
\end{figure}

\begin{figure}[t]
\begin{center}
\begin{subfigmatrix}{1}
\subfigure[]{\includegraphics[width=0.4\linewidth]{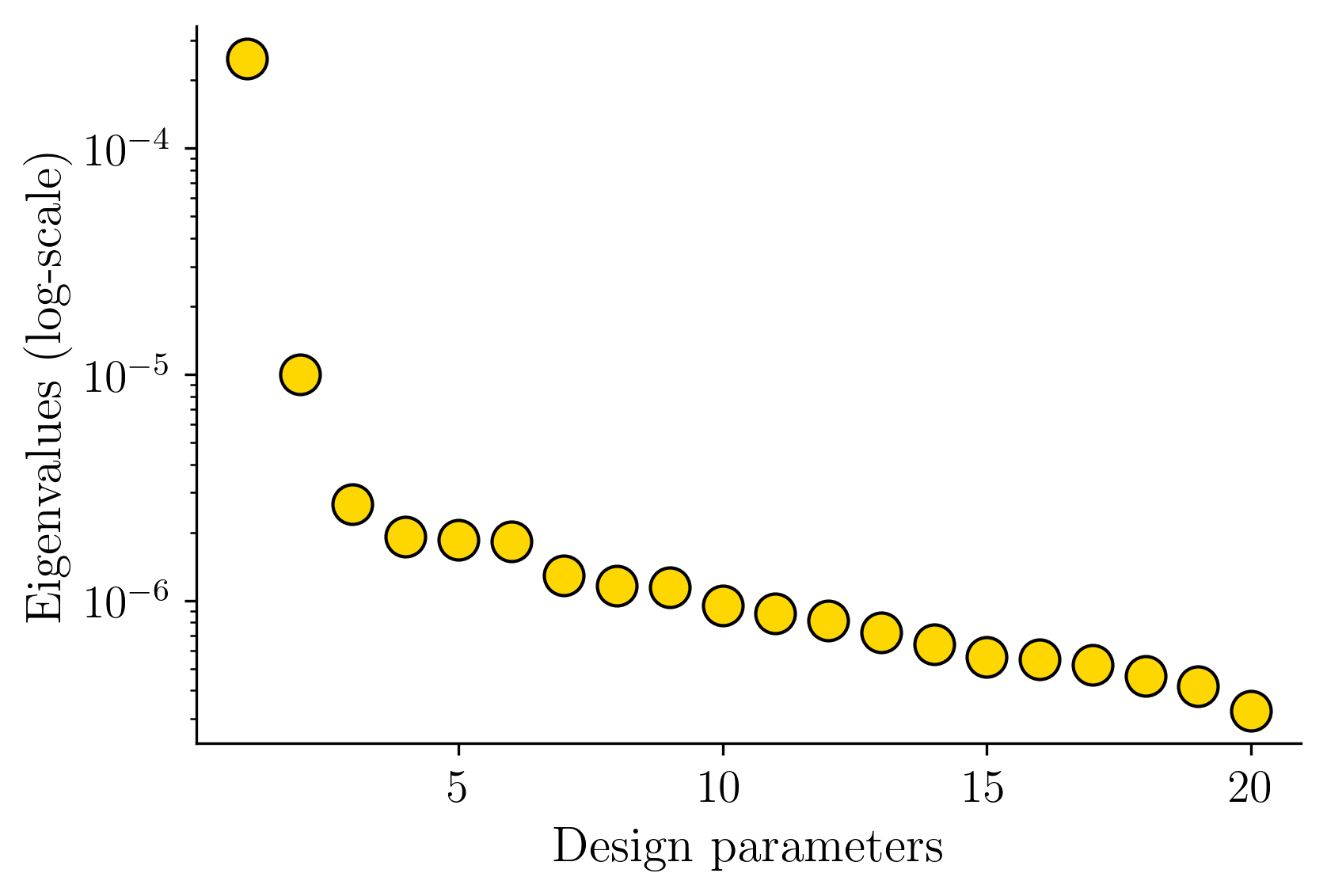}}
\subfigure[]{\includegraphics[width=0.4\linewidth]{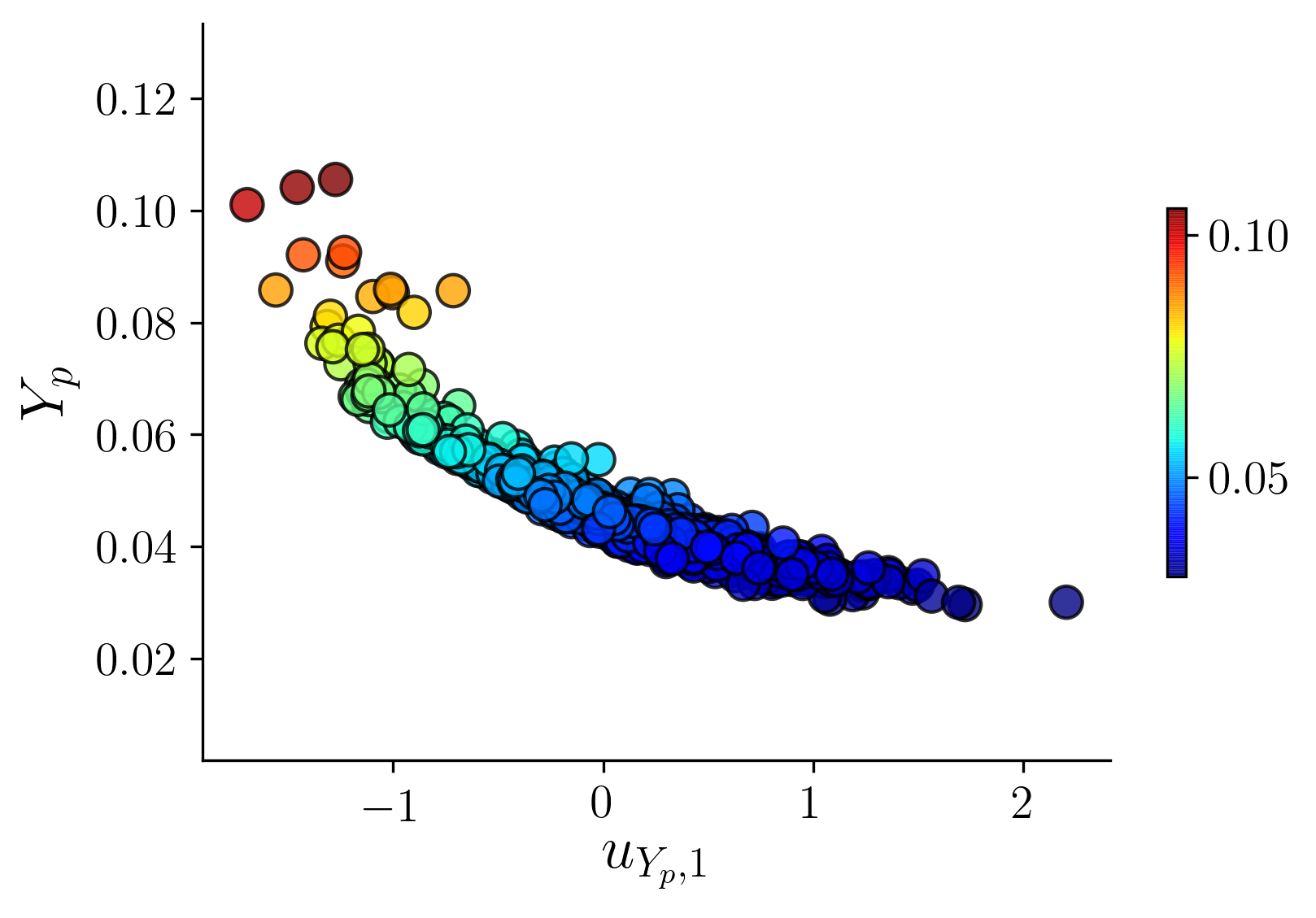}}
\end{subfigmatrix}
\caption{Active subspace computation: (a) eigenvalues of the covariance matrix for loss; (b) sufficient summary plot of CFD values of loss on its first active coordinate.}
\label{fig:active_loss}
\end{center}
\end{figure}

To compute the inactive subspace, we must first find the active subspace. Our DoE comprises 1000 randomly generated designs in $d=20$ which we split into $M=800$ training and 200 testing samples. For each design, we store the CFD yielded loss, mass flow function and isentropic Mach number, although in this paper we focus our attention on the loss only. We fit a global sparse polynomial with order $p=3$ that has a total of $O=1771$ basis terms (see \eqref{equ:total_order}). We use the equadratures package \cite{seshadri2017effective} (see \url{https://equadratures.org}) to construct the sparse polynomial approximation. The resulting coefficients are shown in Figure~\ref{fig:global_poly}(a), where an automated value of $\epsilon=0.001$ was found to be suitable to solve \eqref{equ:cs}. Deploying our model on the testing data-set, we obtain an $R^2$ value of 0.957, which gives us confidence in its predictive capabilities (see Figure~\ref{fig:global_poly}(b)). 
As mentioned in \ref{sec:sparse}, we can use this approximation to estimate the gradients and thus construct the covariance matrix associated with active subspaces for loss $\mC_{Y_p}$. Computing its eigendecomposition yields
\begin{equation}\label{eqn:eigenval_yp}
\mC_{Y_p} = [\mW_{Y_p} \enskip \mV_{Y_p}] \begin{bmatrix} 
\mLambda_{Y_p,1} & \mathbf{0}\\
\mathbf{0} &\mLambda_{Y_p,2}
\end{bmatrix}
\begin{bmatrix}
\mW_{Y_p}^T\\
\mV_{Y_p}^T
\end{bmatrix}.
\end{equation} 
We report the eigenvalues of this covariance matrix in Figure~\ref{fig:active_loss}(a) and, based on this decay, set $\mW_{Y_p} \in \mathbb{R}^{d \times 1}$. In other words, we define our active subspace for loss as being spanned by the first column in $\mW_{Y_p}$, and thus its remaining 19 columns span the inactive subspace for loss. For completeness, we plot the loss for all 1000 geometries on the \emph{active} coordinate of $\mW_{Y_p}$ in Figure~\ref{fig:active_loss}(b). Here we define the active coordinate corresponding to design $\vx$,
\begin{equation}
u_{Y_p,1} = \vx^T  \mW_{Y_p}.
\end{equation}

\subsection{Generating training and testing samples}
\begin{figure}[t]
\begin{center}
\includegraphics[width=0.5\linewidth]{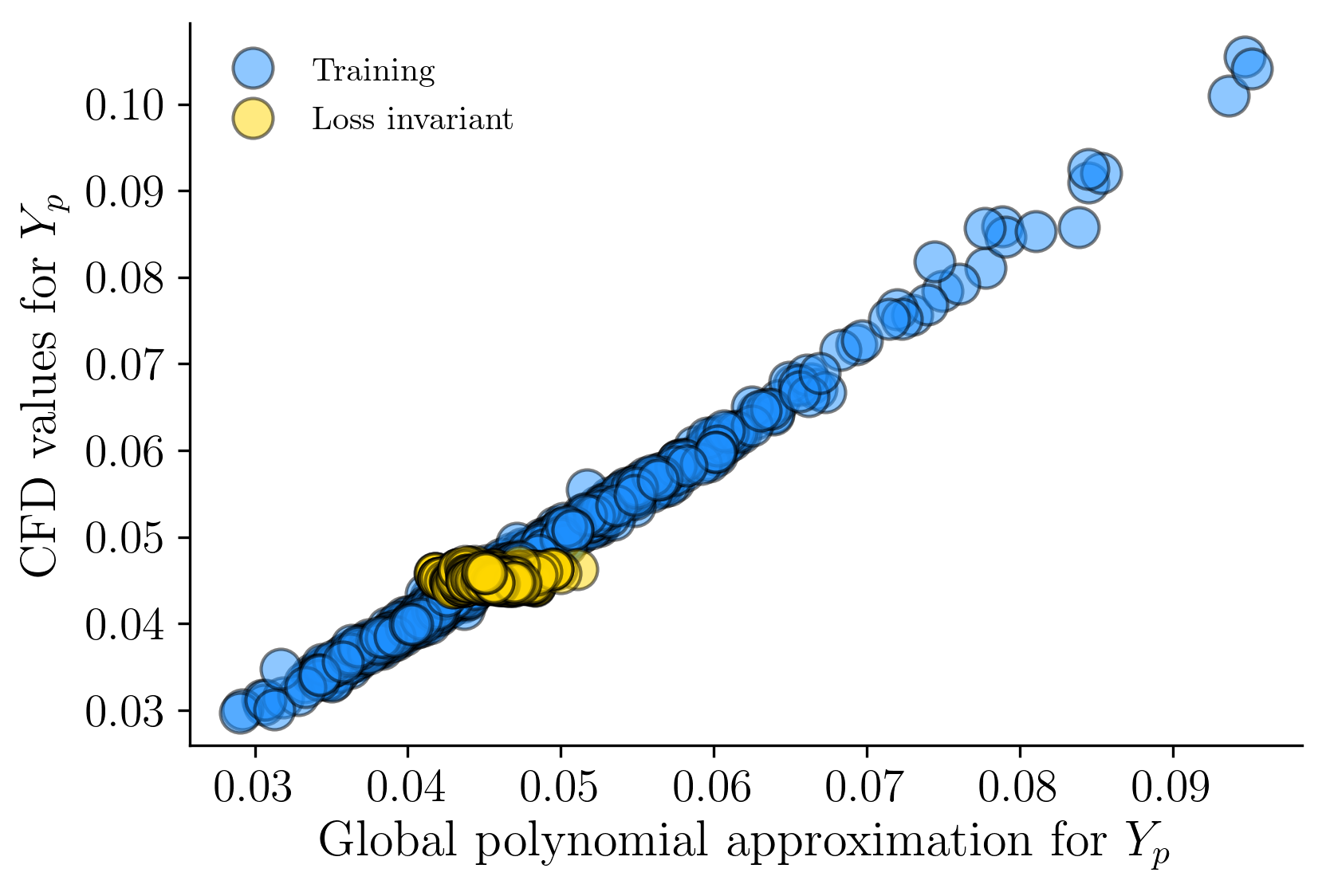}
\caption{Loss invariant designs generated by sampling from the inactive subspace.}
\label{fig:inactive_loss}
\end{center}
\end{figure}

Endowed with the inactive subspace matrix for loss $\mV_{Y_p} \in \mathbb{R}^{d \times 19}$, we can now generate infinitely many blade designs that have approximately the same loss as the nominal LS89. The degree of approximation is dictated entirely by the cut-off value we assume for partitioning the eigenvalues. Here, we generate $H=5000$ new geometries using the hit-and-run sampling strategy detailed in \ref{sec:sample}. We plug in the matrices obtained for $\mW_{Y_p}$  and $\mV_{Y_p}$ into the bounds in \eqref{equ:hitandrunbounds} and set $u= 0$, which corresponds to the \emph{active coordinate} associated with the nominal blade profile (a scalar since the active subspace is 1-dimensional). The values for $\vz$ obtained via hit-and-run sampling are then used to arrive at design vectors $\vx$. To verify that these 5000 additional geometries have similar $Y_p$ losses compared to the datum, we ran 500 of these designs through the CFD solver. The results are reported in Figure~\ref{fig:inactive_loss} and shown alongside the original training data used to fit the sparse polynomial. The length of the two-standard-deviation interval for the $Y_p$ values of invariant samples is 0.0011, much smaller compared to that of random designs which is 0.0185.
This gives us confidence that our methodology has successfully generated designs that have comparable loss values to the LS89. Moreover, it also serves to show that our global sparse polynomial approximation offers an acceptable characterization of the loss.  One important point to emphasize here is that the only computational overhead associated with generating 5000 geometries is the cost of deforming the nominal mesh and generating a new mesh, and thus airfoil coordinates. While it is recommended to run a few CFD evaluations to confirm the veracity of the subspaces if allowed within the computational budget, this verification step is not strictly necessary.

\begin{figure}[t]
\begin{center}
\includegraphics[width=0.5\linewidth]{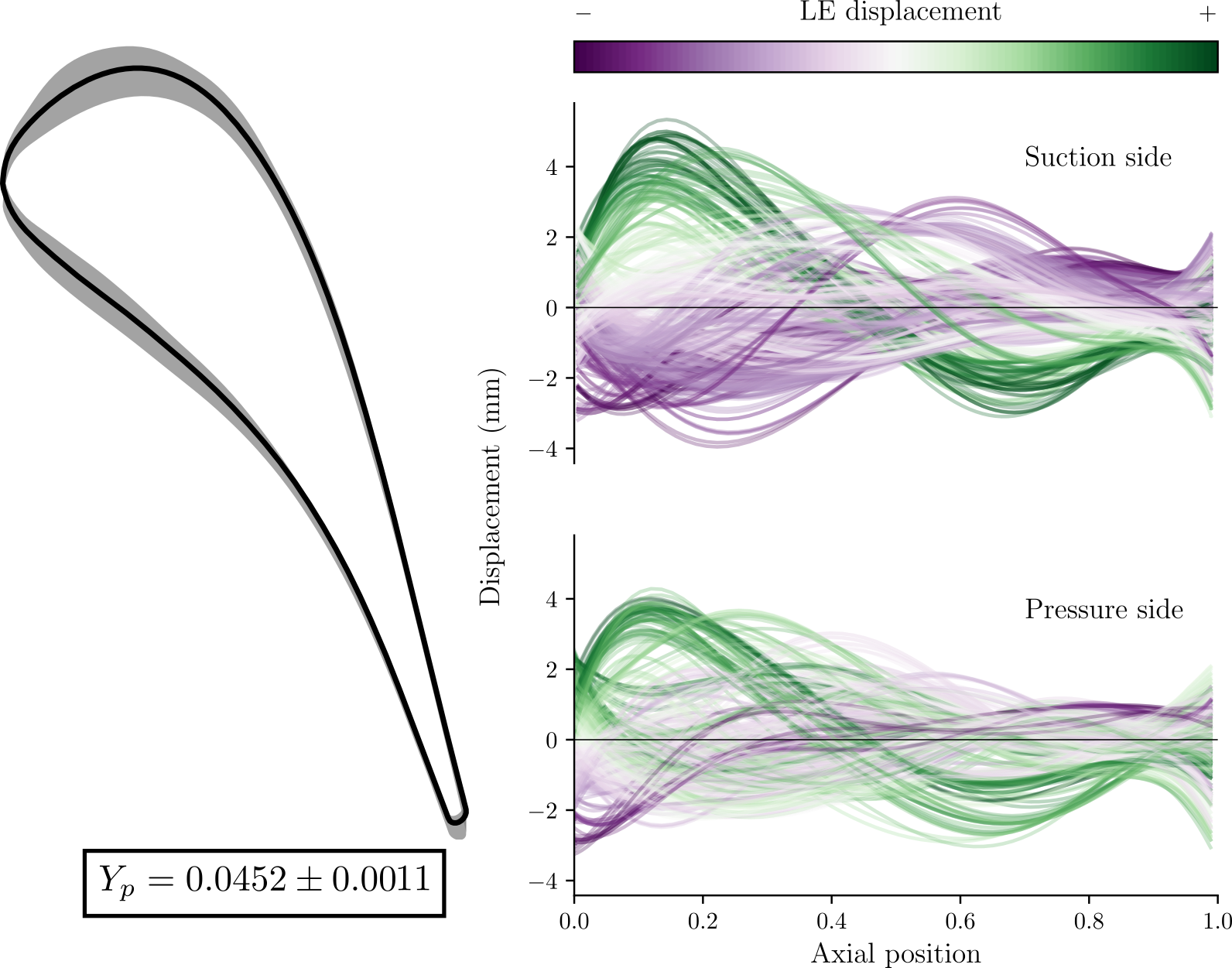}
\caption{The blade envelope for loss.}
\label{fig:be_loss}
\end{center}
\end{figure}

The steps laid out thus far permit us to generate a blade envelope for loss; see Figure~\ref{fig:be_loss}. Any blade that was manufactured within the control zone that satisfies the tolerance covariance (visualized by the LE displacement but quantified by the tolerance covariance matrix $\mS$ \eqref{eqn:tol_cov_matrix}) will admit loss values comparable to the nominal design. 
Note that this envelope has much greater variability compared to the one shown in Figure~\ref{fig:envelope}. This is because in Figure~\ref{fig:envelope}, in addition to loss, the mass flow function is also constrained. We shall study how one generates samples with multiple objectives in the companion Part II paper \cite{wong2020bladeb}. 

\subsection{Use or scrap?}
\begin{figure}[h]
\begin{center}
\includegraphics[width=0.7\linewidth]{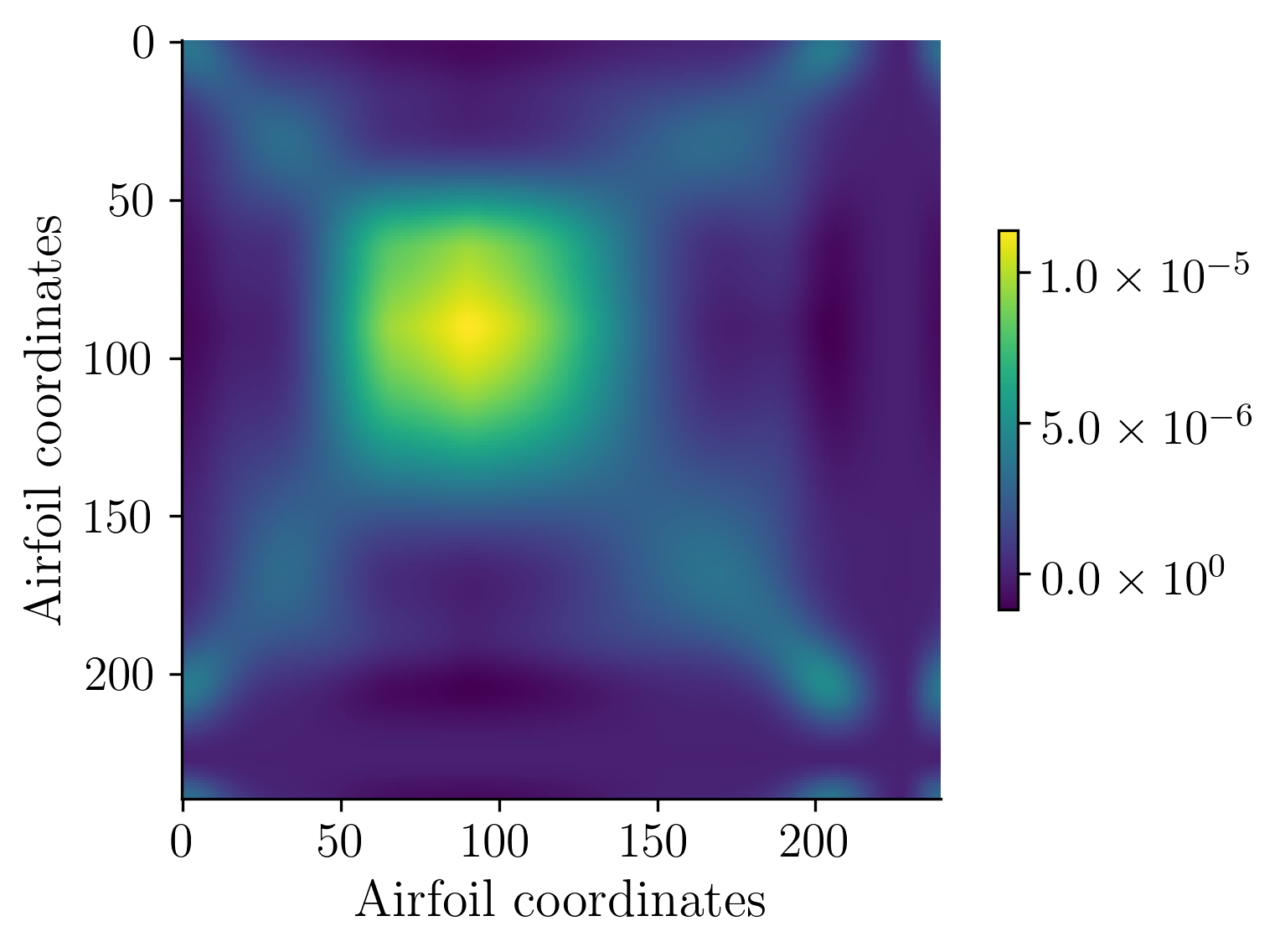}
\caption{The tolerance covariance matrix $\mS$ generated with $H=5000$ inactive samples (left) and with random geometries (right). Colors indicate the value of the matrix at corresponding airfoil coordinates. }
\label{fig:S}
\end{center}
\end{figure}

Following the flowchart in Figure~\ref{fig:schematic}, we compute the ensemble mean $\boldsymbol{\mu}$ and tolerance covariance matrix $\mS$ associated with the different airfoils sampled from the inactive subspace. Our choice of $H=5000$ was dictated by the convergence of the two aforementioned statistical quantities in the $L_2$ norm sense up to 1\% accuracy. The left plot of Figure~\ref{fig:S} shows a heatmap of the tolerance covariance matrix $\mS \in \mathbb{R}^{240 \times 240}$, where the color scale represents the value of each entry in the matrix. To correctly obtain these quantities, each airfoil profile has been carefully normalized to ensure that the horizontal coordinates across the profiles are the same, and thus only the vertical displacements, i.e., thickness / chord, are used in estimating $\boldsymbol{\mu}$ and $\mS$. On the right, we repeat this exercise for random geometries from the original design space. Comparing the two matrices, it can be seen that extra off-diagonal correlation structures are found among the loss-invariant designs. Geometrically, these structures encapsulate the precise curvature requirements for a geometry to be loss-invariant.

\begin{figure}[t]
\begin{center}
\begin{subfigmatrix}{1}
\subfigure[]{\includegraphics[width=0.4\linewidth]{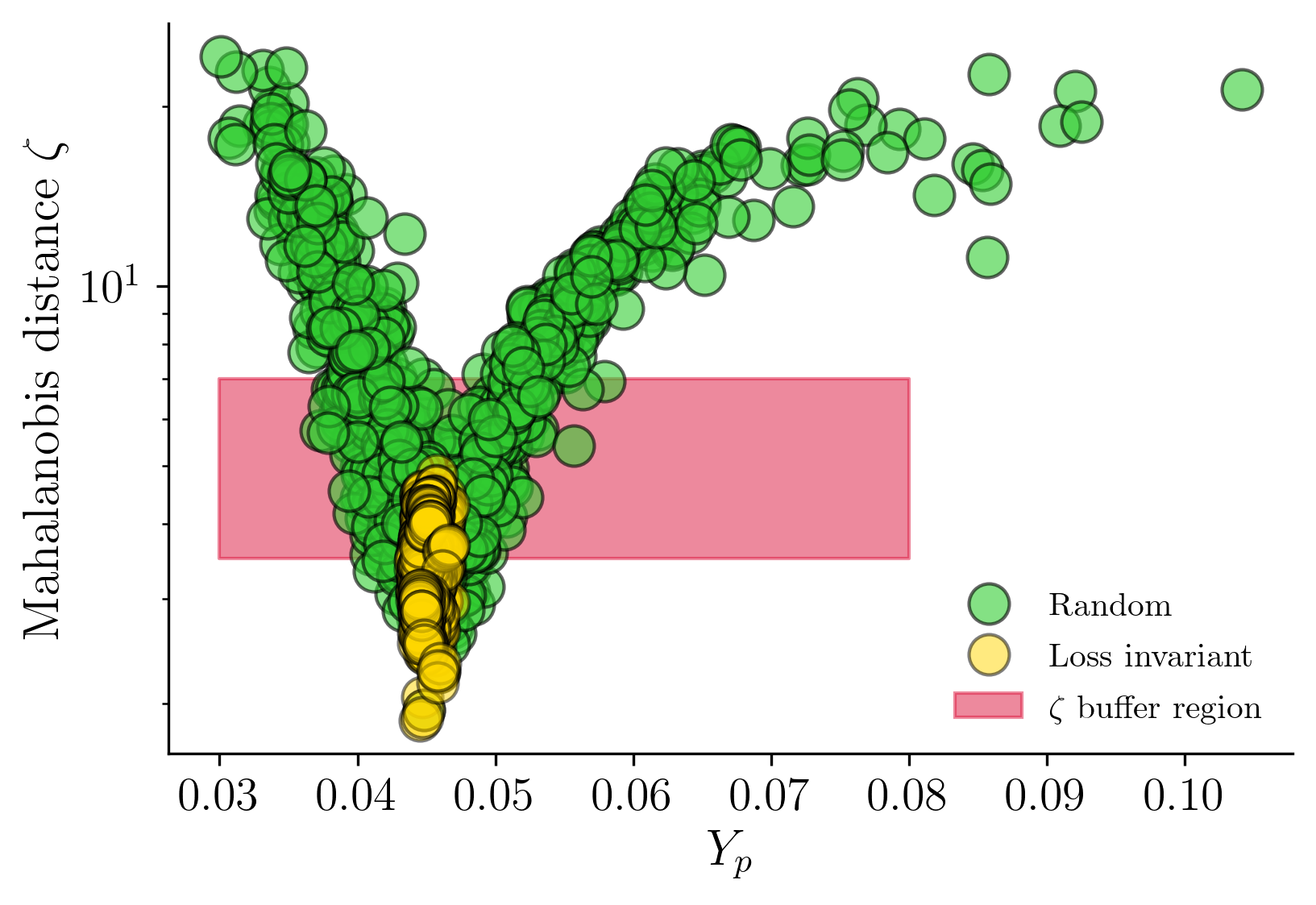}}
\subfigure[]{\includegraphics[width=0.4\linewidth]{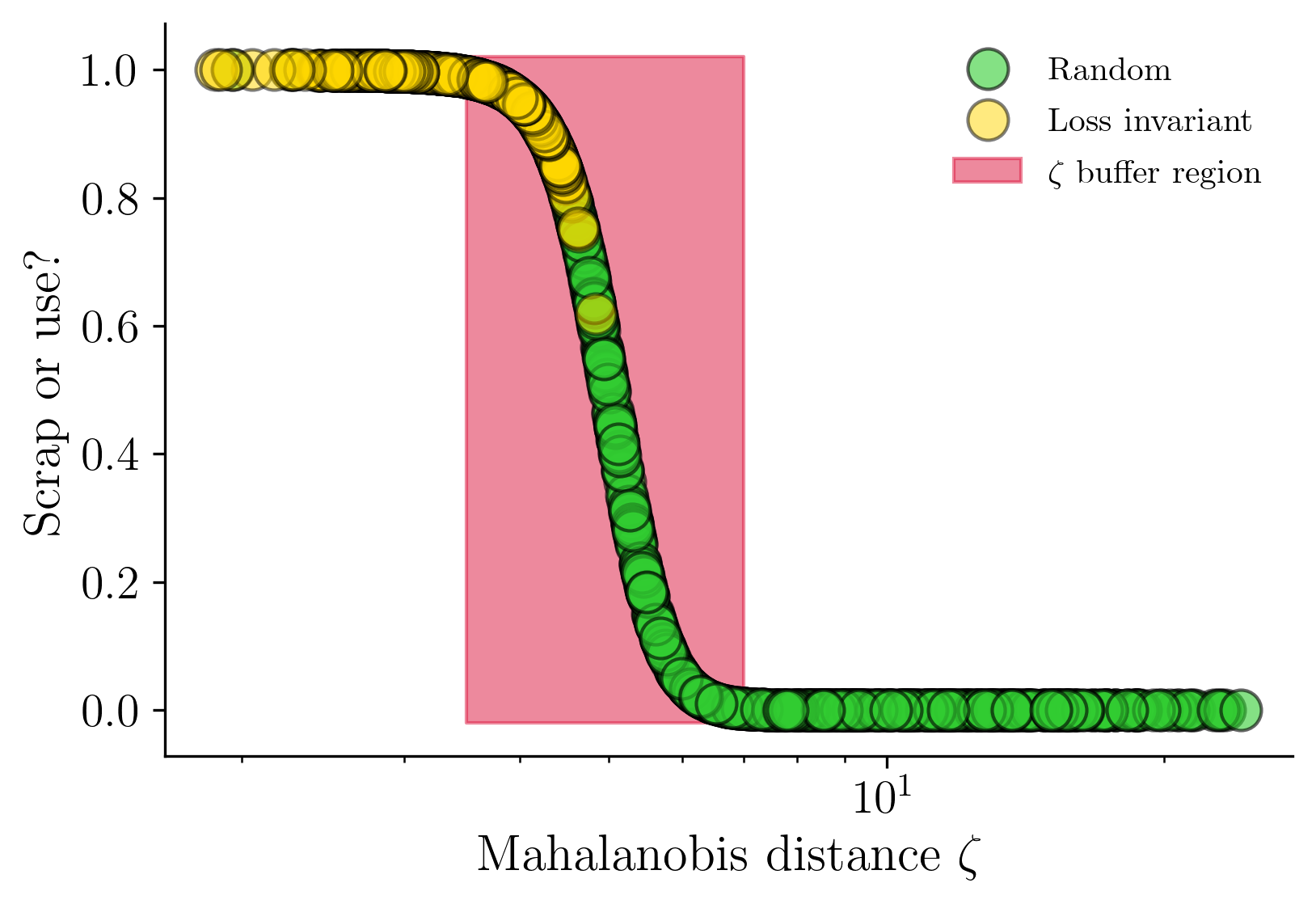}}
\end{subfigmatrix}
\caption{Machine learning blade envelopes: (a) Mahalanobis distances $\zeta$ reported for numerous blades along with their losses; (b) Conversion to a binary scrap-or-use decision where 1 indicates use and 0 indicates scrap.}
\label{fig:useorscrap}
\end{center}
\end{figure}

With these metrics computed, the Mahalanobis distance between any new scanned geometry and the point cloud associated with the inactive samples can be determined. We compute this distance for the 500 out of the 5000 airfoils generated from the inactive subspace and plot them in Figure~\ref{fig:useorscrap}(a). After pruning away the geometries that lie outside of the control zone, we do the same for our original 1000-point training-testing database, for which most of the airfoils admit different values of $Y_p$. A clear divergence in the loss values is observed for $\zeta$ values greater than 7; all the loss invariant designs have low values of $\zeta$, illustrating the utility of our overall methodology. At this point, it is worthwhile to emphasize that this is the result of the particular choice of tolerance covariance, derived from a performance-based analysis of the geometry space using inactive subspaces. To propagate a degree of uncertainty in our assessment, we assume that designs with $\zeta$ values between 3.5 and 7 lie within a $\zeta$ buffer region, also demarcated in the figure.

We use this zone to tune our logistic function and arrive at the following values: $\beta_1=1, \; \beta_2=5$ and $\beta_3=3$; and illustrate the result in Figure~\ref{fig:useorscrap}(b). Here the horizontal axis plots $\zeta$, while the vertical axis represents the binary outcome: a value of 1 implies that the design can be used, while a value of 0 implies that the design should be scrapped. A value in-between---i.e., falling within the buffer region---indicates a geometry that requires further examination before a final scrap-or-use decision. 

\subsection{Evaluating designs from other design spaces}
\begin{figure}
\centering
\includegraphics[width=0.25\linewidth]{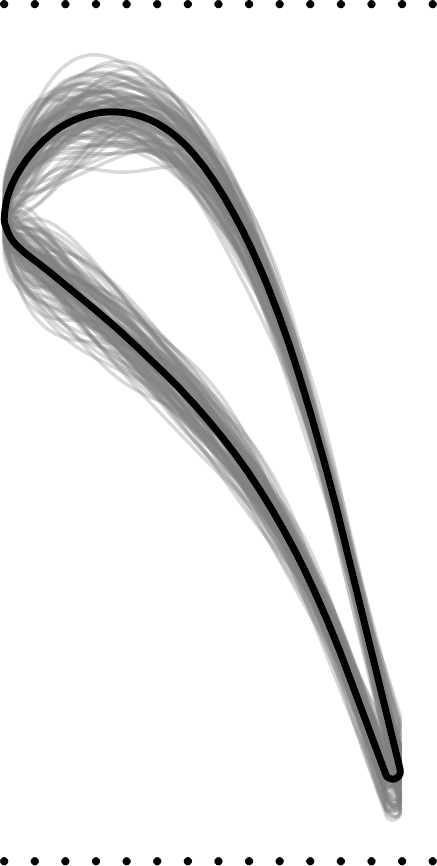}
\caption{Sample designs from the larger design space of FFD design variables. }
\label{fig:ffd_30}
\end{figure}

\begin{figure}
\centering
\includegraphics[width=0.5\linewidth]{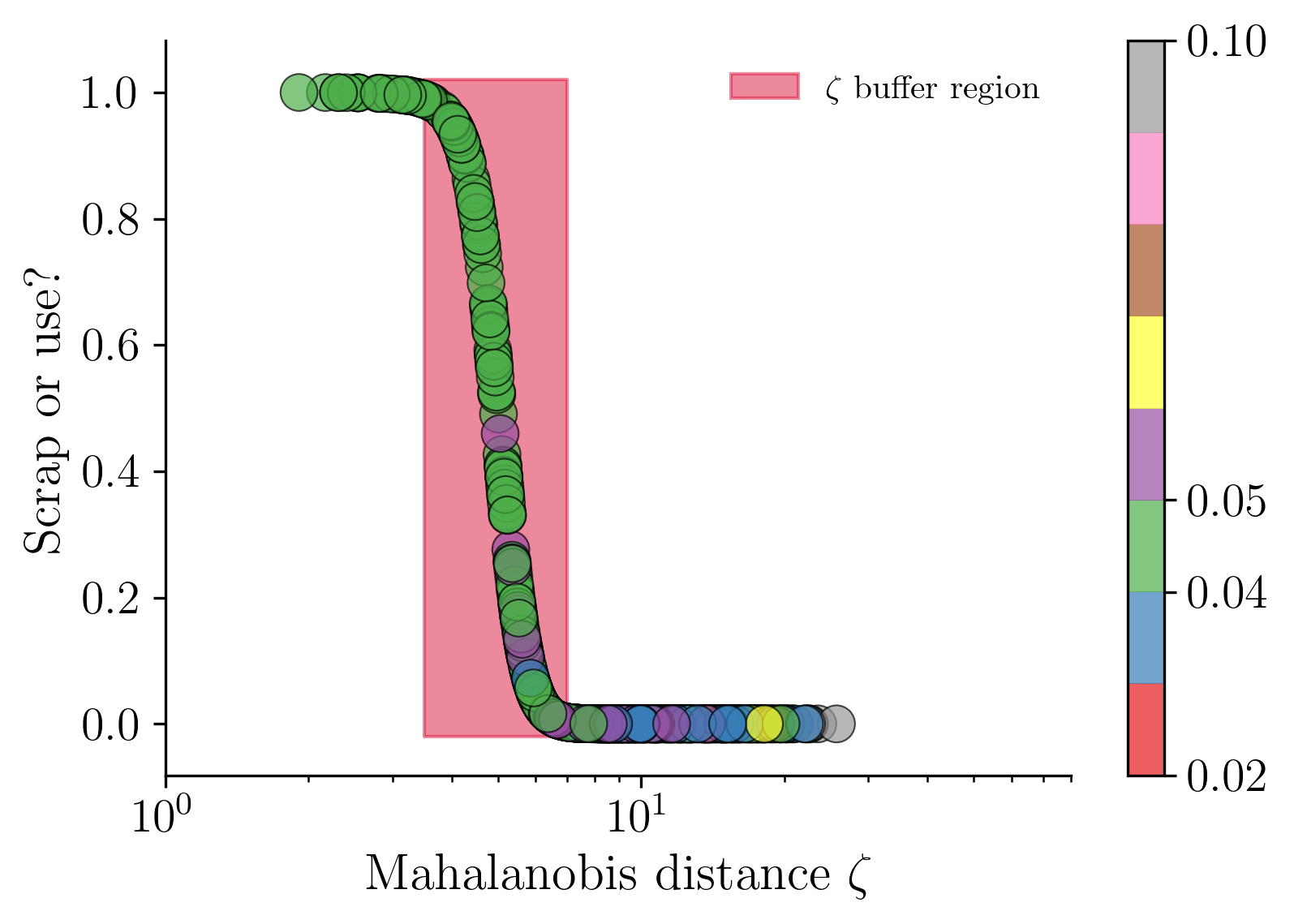}
\caption{Application of the logistic model on profiles generated from the larger design space with $d=30$ design variables. The values on the colorbar are CFD yielded $Y_p$ values.}
\label{fig:useorscrap_30}
\end{figure}

To put our methodology through its paces, we generate 1000 new airfoil profiles from a different design space, one comprising $d=30$ FFD design variables, as shown in Figure~\ref{fig:ffd_30}. We evaluate the loss of these new profiles by running them through the $SU^2$ suite. Then, for each profile that lies in the control zone, we compute $\zeta$ using our previously stored values of $\boldsymbol{\mu}$ and $\mS$---obtained from the 20D design space. These distances are then fed through the aforementioned logistic model and plotted along the vertical axis in Figure~\ref{fig:useorscrap_30}. Here the colors of each marker correspond to the CFD computed loss value. These results attest to the notion that a blade envelope's utility in distinguishing between tolerance-abiding airfoils is not restricted to a particular parameterization. Moreover, the fact that most of the blades with comparable loss values to the LS89 fall either below or within the $\zeta$ buffer region gives us confidence that this workflow can be beneficial and transformative when deployed on real manufactured scans.


\section{CONCLUSIONS}
This paper introduces the concept of blade envelopes and details its underlying methodology. Blade envelopes can be thought of as a blade tolerance draft---a quantitative manufacturing guide---that instructs blade manufacturers and designers what tolerances can be used and whether a used component satisfies these tolerances or not. 

\section*{Acknowledgements}
The first author acknowledges financial support from the Cambridge Trust, Jesus College, Cambridge, and the Data-Centric Engineering programme of The Alan Turing Institute. The second author was funded through a Rolls-Royce research fellowship and the third author was funded through the Digital Twins in Aeronautics grant as part of the Strategic Priorities Fund EP/T001569/1. The fourth author was supported by the Lloyd's Register Foundation Programme on Data Centric Engineering and by The Alan Turing Institute under the EPSRC grant EP/N510129/1. Thanks are due to Bryn Noel Ubald for his help in generating some of the figures, and to R\'{a}ul V\'{a}zquez for his support of this work. We also thank the anonymous reviewers for their comments, which helped improve the quality of the manuscript.

\bibliography{references}
\bibliographystyle{asmems4}

\end{document}